# Performance and User Response of Android's Smartphone-Based Alerts in the 2025 Marmara Ereğlisi Earthquake

S. Mostafa Mousavi[1,2*], Patrick Robertson[3], Richard M. Allen[1,4], Alexei Barski[1], Robert Bosch[1], Nivetha Thiruverahan[1], Youngmin Cho[1], Tajinder Gadh[1], Steve Malkos[1], Boone Spooner[1], Greg Wimpey[1], Edward Shi[1], Marc Stogaitis[1]

[1] Google LLC, Mountain View, CA, USA.

[2] Department of Earth and Planetary Sciences, Harvard University, Cambridge, MA, USA.

[3] Google Germany GmbH, Munich, Germany.

[4] Seismological Laboratory, University of California, Berkeley, Berkeley, CA, USA.

*Corresponding author. Email: mousavim@google.com

## Abstract

This study presents a comprehensive evaluation of Google's Android Earthquake Alert (AEA) system during the Mw 6.2 Marmara Ereğlisi, Türkiye earthquake. AEA detected the event 5.31 seconds after its initiation, alerting over 16 million users. Warning times for weak shaking (MMI III) reached up to 150 seconds, with a median of 56 seconds. While near-source warning windows were shorter, the system achieved 90% true positives and 99% precision overall. The high density of the phone network enabled faster detection than traditional stations, even for this offshore epicenter. Feedback data shows AEA recipients were highly likely to take protective actions, such as "drop, cover, and hold on," or warn others. Timely alerts substantially increased user engagement, perceived usefulness, and future trust. These results highlight how crowd-sourced technology and behavioral insights can effectively enhance seismic resilience on a massive scale.

# Introduction

EEW systems aim to mitigate the impact of earthquakes in urban environments by providing populations with seconds to minutes of warning before the arrival of potentially damaging ground shaking[1-2]. These systems operate by quickly detecting the initial, less destructive seismic waves near the earthquake's origin. They then estimate the potential hazard and rapidly issue alerts to areas that may be affected[3]. EEW systems are based on the principle that electronic signals outpace seismic waves, creating a window for protective measures. This allows for actions such as taking cover, halting industrial operations, or slowing/stopping high-speed transportation. While these measures are intended to lessen casualties and economic damage[4], empirically quantifying these life-saving benefits remains challenging. Furthermore, the potential benefits of EEW must be carefully weighed against the risk of adverse behavioral responses, as alerts themselves can sometimes induce panic or take-action injuries[5].

Traditionally, EEW systems have relied on dense networks of dedicated, high-cost seismic sensors deployed across seismically active regions, such as ShakeAlert in the U.S., the JMA system in Japan, and SASMEX in Mexico[4,6-7]. The substantial infrastructure investment and maintenance required for these systems have limited their global deployment[8]. In response, smartphone-based EEW systems have emerged, leveraging the ubiquitous presence of low-cost accelerometers in modern mobile devices to create distributed seismic sensing networks[9 - 13]. This approach offers a complementary or alternative pathway, potentially providing coverage in areas with sparse sensor networks or augmenting existing infrastructure. Google's AEA system is a prominent example, utilizing billions of Android smartphones to create a global seismic network capable of real-time earthquake signal detection and low-latency alert transmission (milli seconds) via the internet[14]. Encompassing an estimated 70% of the global smartphone market, the Android system can inherently provide an earthquake detection capability wherever human populations reside, spanning both developed and developing nations. The system's efficacy in earthquake detection is contingent upon the spatial distribution of active phones and the influence of human activity induced noise at the time of the seismic event. The AEA system reliably detects earthquakes with magnitudes exceeding M4.5, extending its detection range to several hundred kilometers both on land and offshore. From April 1, 2021, to April 30, 2025, AEA successfully identified a total of 11,231 earthquakes, with magnitudes varying from M1.9 to M7.8 (corresponding to events in Japan and Türkiye, respectively)[14].

On April 23, 2025, at 12:49 local time (09:49 UTC), a strike-slip earthquake of Mw 6.2 struck the central Marmara Sea, 21 km southeast of Marmara Ereğlisi, Türkiye, at a depth of 10 km. The event occurred along the tectonically active North Anatolian Fault Zone (NAFZ), a 1,000 km long fault system delineating the boundary between the Eurasian and Anatolian plates[15-16]. This earthquake was the region's largest in 25 years, generating peak ground intensities of VIII (severe ground shaking) on the Modified Mercalli Intensity (MMI) scale and triggering over 260 aftershocks, including a Mw 5.3 aftershock 13 minutes later, south of Istanbul[17]. Initial reports indicated that buildings in Istanbul (75 km northeast of the epicenter) experienced moderate shaking at MMI V, and at least 359 people were injured in the city, predominantly due to panic rather than direct structural damage[18-19]. While the offshore epicenter mitigated severe structural damage, site amplification in soft sediments led to irregular shaking distribution, with the strongest motions recorded in Çorlu, 43 km from the epicenter[20]. The event highlighted persistent seismic risks in the NAFZ's "seismic gap," where tectonic stress accumulation remains unrelieved since the 18th century[21].

The M6.2 event triggered Google's AEA system, providing a valuable case study for evaluating smartphone-based EEW performance in a real-world scenario. Notably, alerts were delivered to millions of users (90% True Positive, 1% False Positive, and 9% False Negative rates - 99% Precision) in the region, providing more than a minute of warning before the strongest shaking arrivals, highlighting both the potential and limitations of crowd-sourced EEW in densely populated, high seismic risk urban centers. Subsequent statistical analyses were performed to identify relationships among the variables. The user feedback analyses resulted in insightful findings such as the existence of a strong link between user trust and alert timeliness. This study analyzes the AEA system's performance during the event, focusing on detection timing, alert dissemination, timeliness, and user experience, offering insights into the practical role of smartphone-based EEW in mitigating seismic risk.

# Results

## Detection and Strong Motion Arrivals

Türkiye's high seismicity rate makes it one of the top three countries for which AEA has issued alerts. Since AEA's launch in Türkiye in June 2021, more than 116 million alerts have been issued for 203 detected events in the country (as of Sep 16, 2025), including the February 6, 2023, Mw 7.8 Pazarcik and Mw 7.5 Elbistan earthquakes (see Allen et al., 2025 for AEA performance for these events).

Figure 1a shows recorded acceleration data for the Mw 6.2 Marmara Ereğlisi earthquake by triggered Android phones, arranged by their distance to the earthquake epicenter. In this context, a phone is 'triggered' when its on-device detection algorithm identifies a sudden change in the acceleration time series indicative of a P- or S-wave arrival. Upon detecting this brief event, the phone subsequently transmits its continuous acceleration information (the recorded waveforms) and approximate location to the backend server. The recorded waveforms clearly show the moveout of P and S wave energies across the region. The earthquake origin time (09:49:11.9 UTC, based on the USGS estimate) and AEA's detection time (09:49:17.21 UTC) are indicated by solid vertical red and blue lines, respectively. Just 5.31 seconds after the earthquake initiation, the radiated P wave triggered a few hundred phones, leading to the earthquake's detection by the AEA system and the issuance of the first alert 0.43 seconds later, based on an initial magnitude estimate of Mw 4.83. This event was detected and alerted by the Earthquake Network app[10] as well.

As the emitted seismic waves trigger more phones and new data arrives at the AEA server, the estimate of earthquake source parameters, and subsequently the alerted area, by AEA system was updated to reflect the ongoing fault rupture process. During the Ereğlisi earthquake, AEA estimates of magnitude rapidly increased within 15 seconds after the detection reaching the maximum value of 6.06 Mw (Figure S1). This resulted in issuing several alerts delivered to more than 16 million Android users in the region (across Türkiye, Greece, Romania, and Bulgaria). The time and extent are represented by dashed green lines on Figure 1c. The major cities in the region and their distance to the epicenter are presented on the right side of the plot to depict the covered cities by AEA alerts. The spatial dynamics and real-time evolution of these alert polygons are fully animated in Supplementary Movie 1. The median location estimation error of AEA for this earthquake was 3.0 km.

Given the known seismic hazard[22-24], the region is well-instrumented and monitored by Turkish seismic networks. In Figure 1c, seismic waveforms recorded by AFAD seismic stations are plotted similarly as a function of distance to the epicenter (Figure 1c). There are only two seismic stations (stations 3429 and 5906, 25.04 km and 29.15 km away from the epicenter respectively) within 30 km of the epicenter. This is while 560 phones triggered within this distance with 18 phones being 23 km away from the epicenter. We marked the first instance where the peak acceleration value (across three channels) exceeded 6.2% and 0.3% of g (approximating MMI V and III intensity, respectively) on the waveforms. The alerts provided by AEA generally precede the arrival of these ground shaking levels at most locations. Even though this earthquake happened offshore in a generally well-monitored region, phone sensors detected the earthquake signals sooner than conventional seismic stations due to the dense concentration of smartphones immediately along the coast. While this demonstrates the potential of phone-based EEW systems to augment detection in highly populated areas, this performance is heavily contingent on user density. In regions with sparse populations or poor azimuthal coverage, phone networks are unlikely to outpace traditional, well-distributed seismic stations. This observation is accompanied by the ample warning (~1+ min) provided at distant locations.

This demonstrates the potential of phone-based EEW systems. The move out in Figure 1a shows that initially phones were triggered by the earthquake's P wave.

AEA recorded two aftershocks from the Mw 6.2 event, occurring at 09:51:27.3 UTC and 10:02:16.98 UTC. The first aftershock was estimated by AEA as Mw 4.13, which did not trigger an alert. Other organizations reported varying magnitudes for this aftershock: EMSC as M5.3, KOER as M4.7, and AFAD as Mw5.9. One factor contributing to the underestimation of magnitude, as reflected in the number of triggered phones (1,084), could be how people interact with their devices after receiving mass alerts for the mainshock. Furthermore, the brief interval (approximately 2 minutes) between these events, leading to the contamination of the second event's initial arrivals by coda waves from the first, further complicates the accurate magnitude estimation of the first aftershock. The wide range of estimated magnitudes from local seismic networks (AFAD and KOER), coupled with the complete absence of this event from regional and global networks (USGS, INGV, NOA, CENALT), highlights the inherent uncertainty. The second event (10:02:16.98 UTC) was reported as Mw 4.5 by AFAD, but AEA estimated it as Mw3.8, so no alert was issued. A subsequent (to the second aftershock) M5.0 event occurred 22 seconds later, at 10:02:32. Its triggered phones were recorded within the same 1 min time window as the Mw 4.5 event (Figure S2), leading AEA to not identify it as a separate event. Aftershock sequences like this present a major challenge for earthquake early warning (EEW) systems[25].

## Alert Delivery and Warning Time

Figure 2b illustrates that the peak ground acceleration (PGA) appeared on phones a short time after the S wave. The black dots in the plot indicate the device clock time when alerts were displayed on phones, representing the actual warning time users received before the peak ground acceleration (PGA) occurred at their specific location. Warning times for PGA were computed to be over two minutes, and alert delivery delays were found to be mostly <0.1 sec (Figure 2d). The phone data presented is limited to approximately 66,000 alerted and triggered phones, which also participated in the detection process. This represents a small fraction of the total 16 million Android phones that were alerted for this event.

## Alerting Performance

To assess AEA's alerting effectiveness for this event, we first gathered ground truths. This allowed us to quantify the extent of ground shaking at weak (MMI ≥ III) and moderate to strong (MMI ≥ V) intensities. We primarily use the USGS ShakeMap as the ground truth for the ≥ MMI V region. We determined the MMI III level by combining information from several sources: a) intensity contours from USGS ShakeMap, b) intensity contours from INGV ShakeMap, c) felt reports from USGS, d) felt reports from EMSC, and e) an estimated area of felt shaking derived from a Google search survey. In areas where AEA identifies an earthquake, residents searching for earthquake information are prompted to confirm whether they experienced any shaking. Binary information is primarily gathered for detection verification, particularly for events where immediate ground truth data is unavailable. As established in classical macroseismic intensity scales, perception of lower-intensity shaking (e.g., MMI III) is not binary across a population; it is typically felt only by 'some' or a 'few,' while most individuals may not notice it. Because our Google Search survey relies on users actively seeking earthquake information, it inherently samples the motivated subset of the population who perceived the vibration, rather than the total resting population. Nevertheless, there is a strong correlation between the high proportion of "yes" responses to the "did you feel an earthquake" question and the ≥ MMI III region. Out of 239,779 users in the affected region who responded to the survey within five hours of the Mw6.2 earthquake, 231,710 confirmed experiencing the earthquake's shaking. While these respondents may represent a fraction of the total population in distant regions, this massive absolute volume of felt reports validates the widespread demand for situational awareness at these lower intensities. Figure S3 shows the distribution of the search survey results. Areas with high ratio of YES response correlate well with the MMI-3 contour and felt report distributions by USGS and EMSC.

An initial magnitude estimate of Mw 4.83 led to the issuance of the first "Be Aware" alert (for MMI III shaking) 5.31 seconds after the earthquake's origin time. Less than 4 seconds later, the magnitude estimate increased to Mw5.03, prompting the first "Take Action" alert. At both levels, larger alert areas were issued 12.71 seconds after the origin time when the magnitude estimate reached Mw 5.82. The last alert was issued ~ 5 seconds later (17.12 s after the origin time), based on an estimated magnitude of Mw 6.04 (Figure S1). Figure 3 illustrates the first and last alerts, displaying the alerted region, the observed MMI III shaking (ground truth), and the propagation of ground shaking (Shaken Zone) at the time of the alert. We used the group velocities defined in[26] to compute the extent of ground shaking at specific intensity levels at each time point. Green areas represent the areas that received timely alerts at the targeted intensity levels. Almost all "Be Aware" alerts were successfully issued and received before ground shaking began. However, the limited number of "Take Action" alerts issued before the elevated shaking (MMI V) were largely restricted to offshore regions. This spatial underestimation occurred because the AEA system's final magnitude estimate saturated at Mw 6.04, falling short of the actual Mw 6.2 event magnitude (Figure S1). Consequently, the calculated 'Take Action' polygon based on ShakeAlert relations was too small to encompass the full extent of the inland MMI V ground truth. Consequently, inland populations who experienced moderate ground motions (MMI V) did not receive a "Take Action" alert before the arrival of the stronger shaking. They had, however, already received the "Be Aware" warning for the lower-intensity shaking.

Regional spatial grids were classified by alert data (Figures 3e-f and 4). Our evaluation calculates alerting performance at the spatial grid level using total human population data rather than individual device counts. Based on this methodology, the system successfully provided spatial coverage to geographic areas encompassing approximately 36 million residents. Therefore, 90% of the general human population residing within the ground truth shaking region could potentially receive timely and correct alerts, provided they possessed an active, online Android device. Overalerts were minimal (~1% of the population). Most of the population experiencing weak shaking (MMI II-III) without AEA alerts were in northwest Romania. The true positive rate for timely warnings (Be Aware) was 90%, resulting in 99% alerting precision. Be Aware alerts were sent to Istanbul ~ 5 seconds before the shaking. The provided warning time at the outer edge of the true positive region (e.g., Izmir) is more than 70 seconds long. It is important to note that these distant locations are classified as True Positives because our compiled ground truth data (including dense clusters of 'Yes' responses in our felt surveys) confirms the population in these outer edges actively perceived the MMI III shaking. For the True Positive population (the ~36 million individuals in geographic grids that received timely and correct alerts), the median on-time warning was 56 seconds. The widest part of the green distribution in Figure 4c, which represents the mode of warning time distribution, suggests that a substantial portion of the alerted population received well over 60 seconds of warning (with the 75th percentile reaching 73 seconds). Most individuals who received a timely warning were given a substantial amount of warning time, often many seconds, as indicated by the overall data. While False Positives and Late Alerts were minimal, a 9% False Negative rate was observed, indicating that some individuals likely experienced the earthquake shaking despite not receiving AEA alerts. The warning time distribution (Figure 4c-d) indicates that when warnings were timely, they generally provided many seconds of lead time.

“Take Action” alerts resulted mainly in the false negatives (99%). Only 117k (1%) of the ~8M population exposed to the MMI V shaking level could receive “Take Action” alerts 7 to 18 seconds after the arrival of ground shaking (at MMI V intensity level). However, the majority of this population were already notified by “Be Aware” alerts.

## User Feedback Analysis

This section reports on user feedback for the AEA system alert following the M6.2 Marmara earthquake, drawing from two nonrandom, self-selected data sources (meaning participation was voluntary and not based on probability sampling of the general population): 67,056 in-alert user surveys, and 55 supplemental USGS DYFI questionnaires. Our primary goals are to determine the perceived performance and timeliness of the AEA system, characterize the spectrum of user behavior, assess the system's usefulness and user satisfaction, and identify factors that influenced its effectiveness. In the analysis that follows, we will describe each dataset, present our statistical findings, and discuss user perspectives on AEA performance during this specific earthquake.

## AEA User Feedbacks

Accompanying the Android alert notifications for the Marmara Ereğlisi earthquake, a feedback survey posed six questions[14] including, “When did you receive the alert?” There were 67,056 responses to this question. A total of 50.5% of users indicated they received the alert before experiencing the shaking. In contrast, 26.9% received it during the shaking, and 22.6% received it afterward. Please note that this is from the Android users who actually filled the survey and different from Figure 4 which is based on theoretical computations for the total population of the region. While the user feedback data provide real world data, the nuances in perception of different users about shaking and its arrival with respect to the alert might make it less precise than the computed values in the previous section. Although system-side data indicates that alert delivery delays to the devices were mostly under 0.1 seconds (Figure 2d), we cannot rule out that isolated discrepancies may occasionally stem from individual device constraints, such as poor internet connectivity. However, the subjective nature of human perception during a sudden crisis remains the primary factor for the variance in these survey responses. An analysis by distance from the epicenter reveals a clear trend (Figure 5a): while most users within 50 km received the alert during or after the shaking, the proportion of pre-shaking alerts increased rapidly with distance, becoming the dominant experience by a substantial margin for those 80 to 320 km away. These survey results align strongly with the technical alert performance analysis presented in the previous sections. Alert recipients were asked (through the AEA feedback system) to rate the helpfulness of the alert they received with response options in a five-item Likert Scale format ranging from “extremely useful” to “totally useless.” Overall, the alert was well received by respondents with 92.0% rating the alert as extremely or somewhat useful, 4.7% rated it somewhat or totally useless, and the remaining 3.4% had no opinion on usefulness (N = 67,056) (Figure 5).

## Did You Feel It? (DYFI) Survey

The U.S. Geological Survey's "Did You Feel It?" (DYFI) service meticulously gathers macroseismic intensity data from citizen reports, complementing instrumental seismic observations. Recently, a supplementary questionnaire focusing on EEW alerts has been integrated into the DYFI survey, as detailed by Goltz et al. (2023). This questionnaire systematically records user data encompassing their attitudes, actions, and evaluations concerning the alert's efficacy. Specifically, it probes aspects such as the user's circumstances at the moment the alert was received, whether the alert preceded, coincided with, or followed the onset of shaking, actions undertaken in response to pre-shaking alerts, and overall assessments of the alert's utility and preferences for future notifications. 55 responders who reported their felt shaking through USGS DYFI system completed the accompanied supplementary survey on EEW alerts, offering a concise yet valuable independent dataset for the M6.2 event. This survey is substantial because it includes responses from users who did not receive an alert and can correlate these responses with the estimated Modified Mercalli Intensity (MMI) of ground shaking. As noted earlier, this event was also detected and alerted by the Earthquake Network app. However, our hypothesis remains that the EEW alerts reported in this specific survey predominantly originated from AEA. This is based on the much larger scale of alerting of the AEA system and the observation of a similar timing distribution found in the larger AEA user feedback survey, where 47.8% of respondents received the alert before shaking, 30.4% during, and 13.0% after. A critical observation from this dataset reveals a strong correlation between the timeliness of alerts and the intensity of shaking. Alerts received before shaking primarily occurred in areas with lower intensity (averaging MMI 3.7). Conversely, alerts received during or after shaking were located in higher intensity zones (averaging MMI 4.95 and 4.1, respectively). Our prior analyses are supported by this evidence; it is notable that 8.7% of respondents reported not receiving an EEW alert. These responders could be either outside of the AEA's alert contour or not an Android user.

## Variable relationships

Following the methodology of Goltz et al[28], we apply nonparametric statistical methods to analyze social media datasets, evaluating statistical significance with conventional thresholds ($p<0.05$, $p<0.01$, and $p<0.001$). We also used Effect Sizes[29] to reveal the actual size or magnitude of the effect in addition to its significance. The Pearson chi-square test for independence was used to examine the relationship between two categorical variables. This test compares observed values to expected values, assuming no association between the variables. The dependent variables analyzed included the actions taken in response to alerts, respondents' assessment of the usefulness of AEA alerts, and their prospective trust in future alerts, informed by their prior experience.

### Response to the alert

The recipient's response to an alert is crucial as it indicates the alert's effectiveness in encouraging users to act in ways that enhance safety and reduce casualties. To assess this, users were asked, "What did you do after receiving the alert?" and provided seven fixed-choice answers, plus a "None of the above" option. Multiple actions could be selected. We excluded the 5,991 (8.2%) responses of "None of the above" from subsequent analysis. Among the remaining responses, "sharing the alert with others" was the most common action (25.4%), followed by the Drop Cover Hold On (DCHO) action (20.9%). These actions significantly outnumbered "no action" responses (17.7%). Following Goltz et al. (2024), we categorized the responses into three broad classes: active, passive, and no action. As defined by Goltz et al. (2024), 46.4% demonstrated active responses (composed of 25.4% who acted to warn others by sharing the alert, and 20.9% who took physical protective action via DCHO), 36.0% showed passive awareness (e.g., waiting or searching for information), and 17.7% exhibited inaction (Figure 6). A summary of variables pertinent to these responses is provided in Table 1.

People who felt strong shaking were far more likely to take an "Active response" (21,813 observed). Correspondingly, this group was much less likely to be merely "passive aware" or take "no action." People who felt weak shaking are heavily over-represented in the "passive aware" category. This experience validated the alert, making them aware, but wasn't intense enough to force a widespread "active response". People who felt no shaking were vastly more likely to take "no action" (2,224 observed). Logically, this group was also the least likely to have an "active response," since they perceived no physical threat. In short, the stronger the shaking, the more active the response. Weaker or absent shaking correlates strongly with passive awareness or inaction.

Individuals' actions varied significantly based on whether the alert was received before, during, or after the shaking. Both "Active response" and "Passive aware" actions were associated significantly more with alerts that were reported to be received before shaking arrival. "Active response" was also more common than expected during shaking. Conversely, the "No action" category is defined by what didn't happen before the event. Far fewer people than expected chose to do "No action" when the alert arrived before the shaking (4,742 observed). The "No action" group is massively over-represented in the after shaking category (3,179 observed). The alert effectively prompts proactive behavior (both active response and passive awareness) before the shaking begins. The "No action" group is largely composed of people who either decided to do nothing during the event or, looking back after it was over, reported having taken no action.

Receiving an alert before substantial shaking substantially increased the likelihood of an active protective response and considerably reduced the probability of inaction. However, this trend shifted when alerts arrived after the shaking began, resulting in an overwhelming tendency toward inaction, as the window for an active response had closed. Findings from[28] align with the observation that pre-shaking alerts prompted more active responses than post-shaking alerts. However, unlike[28], our study revealed a statistically significant (although with a small magnitude, Cramer's v of 0.074) difference in active responses between those alerted before shaking (53%) and those alerted during shaking (27%), a finding more aligned with anticipated behavioral patterns. Significant differences in responses were observed across all categories, depending on whether individuals received alerts before or after the shaking occurred.

### Reported usefulness of the alert

The timeliness of an AEA alert significantly influenced its perceived utility. "Extremely Helpful" is the only category that is overwhelmingly associated with alerts received before the shaking (23,470 observed). The number of "Extremely Helpful" responses by users who received the alert after the shaking is significantly less (8,498). This strongly suggests that for this group, the alert itself—the simple act of providing an early warning—is what was considered "extremely helpful," regardless of the quake that followed. Every other rating (from "Not helpful at all" straight through "Very Helpful") follows the opposite pattern. These opinions are formed disproportionately after shaking. People wait for the event to conclude, compare the alert to their experience, and then assign a rating. This pattern is most extreme with negative ratings. The "Not helpful at all" opinion was almost never associated with alerts received before the shaking (only 158 observed, 0.2%) but was massively over-represented in alerts that were received after the shaking (1,237 observed). Alerts received before ground shaking were strongly correlated with a positive assessment of their usefulness (54%). The majority of unhelpful votes (~70%) are associated with alerts received during or after shaking. Specifically, among those who received the alert before shaking commenced, 98% rated it as somewhat or extremely useful.

All the lower-to-neutral ratings ("Not helpful at all," "Somehow helpful," and "Helpful") are over-represented in the "Yes, it was strong" group (i.e., strong shaking is linked to lower helpfulness ratings). This suggests that experiencing a strong quake makes people more critical of the alert's performance (perhaps finding it was too late or not precise enough for the intensity). Conversely, all these same negative/neutral ratings ("Not helpful," "Somehow helpful," "Helpful") are under-represented in the "Yes, it was weak" group. This suggests that feeling a weak quake (which validates the alert without causing a major crisis) is a "sweet spot" that is less likely to generate negative feedback. Most surprisingly, the "No, I didn't feel shaking" response is almost statistically irrelevant. For every single helpfulness category, the observed count is nearly identical to the expected count (i.e., no felt shaking has no effect on helpfulness). This indicates that a perceived "false positive" (receiving an alert but feeling no shaking) did not, by itself, cause people to rate the alert as "not helpful" in this dataset.

### User's future trust

Our data indicate that the utility of an alert serves as the primary determinant of user trust; specifically, an alert that provides actionable preparation time is paramount in fostering user confidence, whereas an untimely alert actively erodes it. Alert arrival prior to ground shaking strengthens user trust, whereas alerts received during or after shaking have the opposite effect, significantly reducing a system's credibility. People who decided to "trust the alert less" are overwhelmingly concentrated in the after shaking group.

People who did not feel the shaking were significantly more likely than expected to "trust the alert less" (398 observed) and also more likely to say it "didn't affect" their trust (1022 observed). Logically, receiving an alert for an event they couldn't perceive (a perceived "false positive") damages credibility. Conversely, people who felt weak shaking were less likely to "trust the alert less" (545 observed). This suggests that feeling *any* shaking, even if weak, acts as a positive validation of the alert and prevents the loss of trust seen in the "no shaking" group. Interestingly, our results suggest that user's trust opinions are simply proportional to the overall sample and aren't uniquely skewed by the *strength* of the quake itself (likely because their opinion is already better explained by the "helpfulness" variable).

# Discussion

Our understanding of how humans respond to EEW systems is still developing, often based on anecdotal evidence, making a comprehensive analysis of their effects difficult[30 - 34]. This study analyzes the performance of Google's AEA system during the Mw 6.2 Marmara Ereğlisi, Türkiye earthquake. Our evaluation, employing a suite of utility-based metrics and drawing upon diverse data sources including phone triggering data, seismic station data, and felt reports from Google Search, USGS DYFI, and EMSC, demonstrates the high precision and substantial warning times delivered by the AEA (Be Aware) alerts. A key finding was the earlier detection of seismic signals by smartphone accelerometers compared to the wavefield's arrival at local seismic networks in this specific geography. This observation highlights the practical value of unconventional seismic sensing networks, such as phone-based systems, in real-world operations. While crowd-sourced EEW systems like AEA inherently offer a significant advantage by extending coverage to high seismic risk regions with sparse traditional sensor infrastructure, the results of this study further suggest a potential for even greater utility. Specifically, our findings advocate for the exploration of hybrid operational models that strategically integrate phone-based detections with conventional seismic networks, thereby leveraging the inherent strengths of both approaches to potentially enhance early detection, refine location and magnitude estimation accuracy, and improve alert delivery.

Analysis of AEA feedback surveys, conducted to understand user experiences with EEW alerts, revealed a significant finding: Android users exhibited a considerably higher rate of active responses (46.6%) compared to inactions (17.7%) after receiving AEA alerts. This observation contradicts previous studies, such as Goltz et al.[28] on the 2023 Mw 5.1 Ojai, California earthquake; Nakayachi et al.[35] on the 2018 M4.4 and M5.9 earthquakes in Gunma and Chiba, Japan; and Vinnell et al.[36] on the 2021 M5.3 and M5.9 New Zealand North Island earthquakes. These earlier studies commonly reported a greater percentage of infections among responders, reaching up to 54% in Nakayachi et al.[35]. Several factors, including the earthquake's larger magnitude, historical context, and cultural variations, may account for the increased percentage of active responders in the current study.

To investigate this further, we compared the ratio of user responses for the M6.2 event with those from all alerted events in Türkiye, Philippines, Chile, USA, and the rest of the world by AEA system (Figure 7). Our results suggest that this event was associated with a higher ratio of active responses and passive awareness than the average. This could be attributed to the specification of this event and its historical context. Moreover, recent catastrophic earthquakes in the country (i.e., 2023 M7.8 Kahramanmaras earthquake) could play a role in elevating the public awareness. This geographical and historical context inherently elevates public sensitivity and tolerance to seismic hazards and, consequently, to the perceived efficacy of mitigation tools like EEW systems. The responses of alert recipients during such critical periods represent a pivotal measure of an EEW system's real-world utility.

Figure 7a indicates that Türkiye and the Philippines show the highest proportions of active response to alerted events (42.3% and 46.8%, respectively). This suggests either successful public education campaigns (e.g., "drop, cover, and hold on") or a population accustomed to taking physical safety measures during earthquakes. Both countries have experienced large ( > M7.0) alerted earthquakes detected by AEA. The Philippines also exhibits the highest rate of AEA-detected events, with minimal overestimation/over-alerting of magnitude (most events are >= M4.5). In contrast, the USA and Chile demonstrate lower rates of active response and a greater inclination towards passive awareness, implying that people are more likely to pause, seek online information, or check on others rather than performing a prescribed physical action. The rate of "no action" in the USA (15.1%) is similar to the average across other regions (14.7%), while Chile reports the highest percentage of "no action" (24.3%) among the listed countries. The Philippines has the lowest rate of "no action" responses at 8.1%. Except for Chile, "no action" represents the smallest portion of responses from Android users receiving AEA alerts. Passive awareness is the predominant action for users in the USA, Chile, and the aggregation of alerts for other countries. For alerted events in Türkiye and the Philippines, "active responses" constitute the majority of actions. This trend (active response > passive aware > no action) is even more pronounced for the 2025 M6.2 Marmara event. The countries with higher ratio of “active responses” (i.e., Türkiye and Philippines) have higher rate of user trust than those with less active response rates (i.e., USA and Chile). Chile has the lowest rate of before-shaking alerts (36.3%) among the listed countries. While this might explain its lowest rate of active responses (23.3%), it is interesting that user’s trust in the system is still relatively high (74.4%). The relative ratio of active responses between Chile, Türkiye and the Philippines (23.3%, 42.3%, and 46.8% respectively) is proportional to the relative ratio of alert timeliness (36.3%, 40.6%, and 47.6% before shaking alert respectively) and the user trust (74.4%, 77.9%, and 86.4% will trust more respectively). It is interesting that the user trust (i.e., will trust more ratio) is the lowest among US responders while the highest rate of before shaking alerts were delivered in this country. However, it is important to note that the alerted events in the US are based on the detection and estimates of USGS ShakeAlert and AEA only delivers ShakeAlert alerts to the Android users through its delivery mechanism.

Contrary to findings by Nakayachi et al.[35], who suggested mobile phone alert recipients prioritize information-seeking, our observations showed a lower frequency of this behavior (10.3% for M6.2 and 20-25% across all events in individual countries). This difference might be attributed to regional variations in earthquake frequency or the information rich delivery of AEA alert messages. A study by Vinnell et al.[36] on responses to Google's Android Earthquake Alert (AEA) system for the 2021 M5.3 and M5.9 New Zealand North Island earthquakes revealed that individuals primarily "stopped and waited" or "sought additional information," with few taking protective actions. The study also highlighted a general lack of public awareness regarding Google's AEA system. However, it's important to consider that these New Zealand earthquakes occurred shortly after the AEA system's launch, when public understanding was likely still developing. Additionally, the earthquake magnitudes and probable ground intensities differed. Despite these factors, user assessments of alert usefulness were largely positive, with over 80% in both events rating alerts as "somewhat" or "extremely useful" due to the mental preparation time and immediate event information provided.

Initial reports indicated that many of the injuries during this event were due to panic rather than direct structural damage. While the survey data in this study cannot definitively measure psychological panic, the high prevalence of prescribed physical actions (e.g., Drop, Cover, and Hold On) suggests the alerts often prompted constructive behavior. Further interdisciplinary research is necessary to fully quantify how early warnings modulate public panic during seismic events. If alerts trigger maladaptive behaviors, such as attempting to exit a building during active shaking, they could potentially lead to worse outcomes. Therefore, as highlighted by McBride et al.[38 - 39], ongoing public education is absolutely essential to ensure alerts consistently trigger safe, prescribed actions rather than panic.Our continuous focus is on enhancing the AEA system by improving its technical capabilities for handling intense aftershocks, refining magnitude estimation for complex seismic events, and minimizing alert delivery latency. Furthermore, research is needed to understand how user responses to AEA alerts vary across regions and over time.

While the performance of the AEA system during the M6.2 Marmara Ereğlisi earthquake demonstrates the potential utility of smartphone-based EEW, it is critical to acknowledge the inherent selection bias in analyzing a single, highly successful event. As noted in EEW literature, an M6.2 earthquake represents a relative 'sweet spot' for early warning systems; it is large enough to generate widespread alerts but avoids the magnitude saturation issues that complicate rapid parameter estimation for M7+ events (Allen et al., 2025). Furthermore, while this specific event yielded a high True Positive rate and minimal false alerts, the global operational reality of the AEA system over the past several years naturally includes instances of missed events, false alerts, and magnitude mischaracterizations in various tectonic settings. A comprehensive, global evaluation of the system's technical limitations across more than 11,000 earthquakes is detailed in Allen et al.[14]. The primary value of the present case study lies not in proving flawless technical performance, but in utilizing the exceptionally large volume of user feedback generated by this specific event to understand human behavioral responses and the dynamics of user trust.

The necessity of distinguishing between these two alerting tiers is especially apparent when considering different operational scenarios. Because 'Be Aware' alerts function as standard notifications that respect 'Do Not Disturb' settings, their massive dissemination during this daytime event yielded high user engagement. However, had this earthquake occurred at night, these notifications would have largely failed to wake a resting population. In such a nocturnal scenario, only the breakthrough 'Take Action' alerts would successfully prompt protective awareness. Given that the 'Take Action' alerting region for this event underestimated the actual extent of MMI V shaking due to magnitude underestimation, framing the system's success primarily around the volume of 'Be Aware' alerts obscures critical performance gaps in high-stakes, real-world conditions.

# Methods

## Detection and alerting

Smartphones, equipped with three-axis accelerometers originally designed for screen orientation, effectively capture ground shaking during seismic events. When plugged in and stationary, these phones report their availability ("online") for monitoring by the Android Earthquake Alert (AEA) system. An on-phone detection algorithm analyzes acceleration time series for sudden changes indicative of P- or S-wave arrivals. Upon detecting a potential event ("triggered"), acceleration information and an approximate location (location coarsened to preserve privacy) is transmitted to the backend server. This detection capability, a core component of Google Play Services, is available on the vast majority of Android smartphones, requiring no application installation.

The backend servers subsequently correlate the pattern of phone activations with potential seismic sources within the spatio-temporal domain. An earthquake is officially declared, and its source parameters—such as magnitude, hypocenter, and origin time—are estimated when a candidate seismic source demonstrates a sufficiently high confidence level in satisfying the observed data. Upon earthquake detection, the system estimates the intensity of ground shaking and its potential aerial extent. For events with an estimated magnitude surpassing Mw 4.5, AEA issues two distinct alert types to users within the projected impact zone: "Take Action" alerts are disseminated to regions anticipated to experience moderate or greater shaking (i.e., ≥ MMI 5), while "Be Aware" alerts are dispatched to areas expected to encounter weak shaking (i.e., MMI 3 or 4). The "Take Action" alert commandeers the entire phone screen, overriding any do-not-disturb settings, and emits a distinctive, attention-commanding sound. Conversely, the "Be Aware" alert manifests as a standard notification, akin to other phone or application notifications, but is accompanied by a characteristic auditory cue. Once the shaking subsides, or if the alert is received post-shaking, these alerts are supplanted by an "Earthquake Occurred" notification. The alerts delivered to Android phone users comprise a concise summary of the event's attributes, precautionary instructions, earthquake safety information, and a brief user survey for feedback regarding alert delivery (further details are available in Allen et al.[14]).

Safeguards were implemented to mitigate false positives, with mechanisms for their swift identification should they arise. The validation of system detections is primarily achieved through two avenues: a) external earthquake catalogs compiled by regional and international bodies, and b) felt shaking reports amassed via Google Search.

## Alert performance evaluation

For the evaluation of alerting performance, the region was discretized into spatial units approximately 20 km in extent (324.29 km²). The eligible measurement space, defined as the cumulative area encompassed by both the alerted and targeted regions at each Modified Mercalli Intensity (MMI) level, was subsequently categorized into mutually exclusive classifications:

a) Correct-AND-Useful [True Positive, TP]: If a grid within the ground truth received an alert before shaking (arrival of S wave) arrives at its center.

b) Correct-BUT-NotUseful [Late Alert, LA]: If a grid within the ground truth (targeted area for alerting at a specific intensity level) received the alert after the shaking arrived.

c) Missed-Alert [False Negatives, FN]: If a grid within the target MMI area (i.e., ground truth) did not receive an alert.

d) False-Alert [False Positives, FP]: If an alerted grid is outside of the target MMI areas.

The fundamental objective of evaluating alerts is to ascertain their real-world efficacy for stationary populations concerning the seismic source. Consequently, the evaluation should prioritize spatial grids encompassing human populations, utilizing the human population (not phone counts) at the grid level as the unit of computation. To determine the warning time, we compute the interval between the alert (specifically, the earliest alert covering the grid) and the arrival of the S-wave at the grid's center.

Based on this classification and population count in each grid, we compute the following evaluation metrics:

$$\text{True Positive Rate} = \frac{TP}{TP + LA + FN} \tag{1}$$

$$\text{Late Alert Rate} = \frac{LA}{TP + LA + FN} \tag{2}$$

$$\text{False Negative Rate} = = \frac{FN}{TP + LA + FN} \tag{3}$$

$$\text{False Positive Rate} = \frac{FP}{TP + LA + FP} \tag{4}$$

$$\text{Precision} = \frac{TP}{TP + LA + FP} \tag{5}$$

## Ethics Statement and Data Usage

This study analyzes two types of user-generated data. First, macroseismic intensity data were obtained from the United States Geological Survey's "Did You Feel It?" (DYFI) system; this constitutes a secondary analysis of publicly available government data. Second, the study analyzes user feedback from the Android Earthquake Alerts (AEA) system. This data was collected by Google LLC as part of standard product quality assurance and system monitoring. The feedback mechanism is voluntary, and data collection is governed by Google's Terms of Service and Privacy Policy. All AEA data analyzed in this study were retrospectively accessed in a fully anonymized and aggregated format. As the study involves the analysis of existing, anonymized administrative and product data, specific Institutional Review Board (IRB) approval was not required. No researchers conducted physical field studies or direct in-person interactions with participants in Türkiye; all data were collected digitally via the global Android platform.

Base maps for spatial visualizations were generated programmatically in Python using the Cartopy library (version 0.25.0). Map tile imagery was sourced from Google Maps via Cartopy's Google Web Map Tile Service (`GoogleWTS`) interface.

# Data availability

Intensity contours used as ground truth have been obtained from USGS's ShakeMap: https://earthquake.usgs.gov/earthquakes/eventpage/us7000pufs/shakemap/intensity. Ground truth felt reports were extracted from USGS Did You Feel It (https://earthquake.usgs.gov/earthquakes/eventpage/us6000qhiq/dyfi/responses) and EMSC (https://www.emsc-csem.org/Earthquake_information/earthquake.php?id=1798415). Station seismic data used in this study are available from Turkey's nationwide earthquake monitoring system, operated by the Disaster and Emergency Management Authority (AFAD) (https://tadas.afad.gov.tr/). Maps and figures in this paper were generated using the Generic Mapping Tools and Matplotlib. User feedback data from the Android Earthquake Alerts (AEA) system was collected as part of standard product quality assurance, is governed by Google's Terms of Service and Privacy Policy, and is thus exempt from specific Institutional Review Board (IRB) approval because it involves the analysis of existing, fully anonymized, and aggregated administrative data. The Android Earthquake Alert (AEA) system data, including phone triggering records and user feedback surveys, are proprietary to Google LLC. Due to strict user privacy protections and governance under Google's Terms of Service and Privacy Policy, these specific datasets cannot be made publicly available.

# Code availability

Our source code used for processing of station data is available at https://github.com/smousavi05/turkey_2025_m6_2.

## Acknowledgements

We would like to thank Brian Williams, Mohammad Khidar, Ed Fernandez, for insightful remarks and suggestions. We are grateful to Dr. David Wald, Dr. Jessie Saunders, and Prof. Francesco Finazzi for the detailed review and insightful comments. We are particularly thankful to David Wald for providing supplementary DYFI survey results for this event.

## Funding Statement

This work was funded by Google LLC.

## Author Contributions Statement

SMM: Conceptualization, Methodology, Validation, Visualization, Writing— original draft; PR: Conceptualization, Supervision, Validation, Writing —review and editing; RMA: Visualization, Validation, Writing —review and editing; AB: Data curation, Supervision,Validation, Writing—review and editing; RB: Data curation, Validation, Writing—review and editing; NT: Data curation, Validation, Writing—review and editing; YC: Data curation, Validation, Writing—review and editing; TG: Data curation, Supervision; SM: Project Administration, Supervision; BS: Project Administration, Supervision; GW: Project Administration, Validation; ES: Project Administration, Validation; MS: Conceptualization, Supervision, Project Administration, Writing—review and editing.

## Competing Interests

RMA is the UC Berkeley PI on the USGS ShakeAlert project and the California Earthquake Early Warning System (CEEWS) project and sits on various advisory committees related to both. The remaining authors declare no competing interests.

TABLE 1: Statistically significant variable relationships (one-sided).

| Variables | Chi-square Statistic | DF | p value | Cramer's V (Effect Size) | Relationship | Total observations (N) |
|---|---|---|---|---|---|---|
| Shaking experience & Active alert response | 1377.0599 | 4 | 0.0000 | 0.0973 | The intensity of the shaking directly predicts the person's level of action. | 72,695 |
| Alert arrival wrt shaking & Active alert response | 708.6294 | 4 | 0.0000 | 0.0743 | People were significantly more likely to take an active response upon receiving an AEA alert before the shaking began. | 64,101 |
| Alert arrival wrt shaking & Alert usefulness. | 4557.3368 | 8 | 0.0000 | 0.2025 | There is a very strong association between receiving an alert before shaking and perceiving it as useful. | 55,590 |
| Shaking experience & Alert usefulness. | 284.2880 | 8 | 0.0000 | 0.0457 | The criticism (lower helpfulness ratings) comes from those who experienced strong shaking, not from those who experienced nothing. | 68,057 |
| Helpfulness & User's future trust. | 9440.6073 | 8 | 0.0000 | 0.3076 | Perceived helpfulness directly predicts the user's future trust. | 49,881 |
| Alert arrival wrt shaking & User's future trust. | 3930.3674 | 4 | 0.0000 | 0.2122 | Receiving an alert before the shaking is the single most powerful factor in building user trust. | 43,662 |
| Shaking experience & User's future trust. | 182.0822 | 4 | 0.0000 | 0.0427 | No felt shaking erodes trust. | 49,889 |

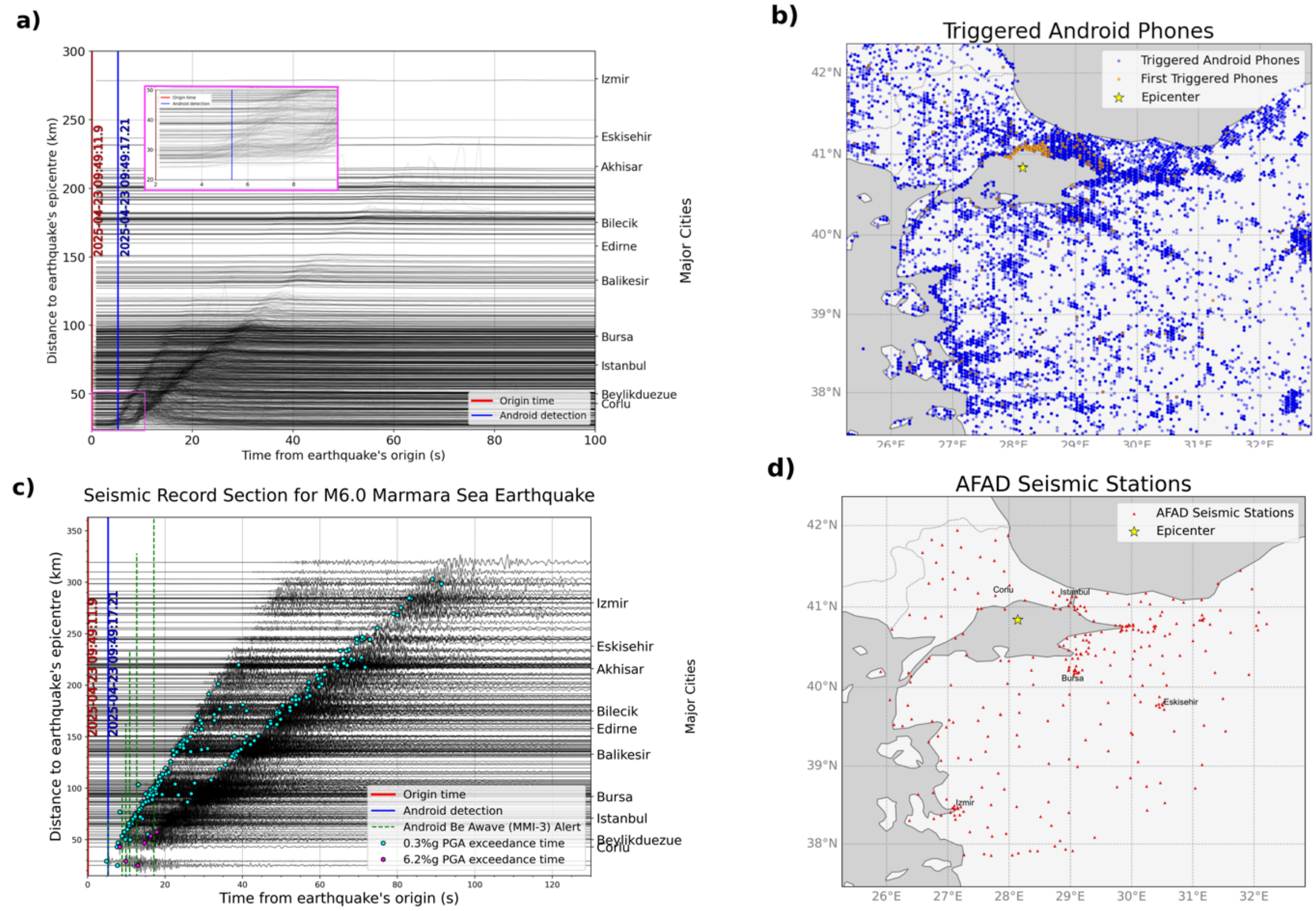

Figure 1: **Comparison of Smartphone and Traditional Seismic Network Detections. (a, c)** Ground acceleration time series from **(a)** triggered Android phones and **(c)** AFAD seismic network recordings (vertical component). Seismograms are ordered by distance, displaying P and S wave moveout. Origin time (red) and AEA detection time (blue) are marked. Dashed green lines indicate the radius of issued AEA "Be Aware" alerts. Waveforms also mark MMI V (6.2% g) and MMI III (0.3% g) thresholds. Epicentral distances for major cities are shown on the right. **(b)** Regional distribution of the ~66,000 online Android phones used for event characterization, with 430 initial P-wave triggers highlighted. **(d)** Distribution of AFAD seismic stations. Yellow stars denote the Mw 6.2 epicenter. Maps generated using Cartopy (v0.25.0); Map data ©2024 Google.

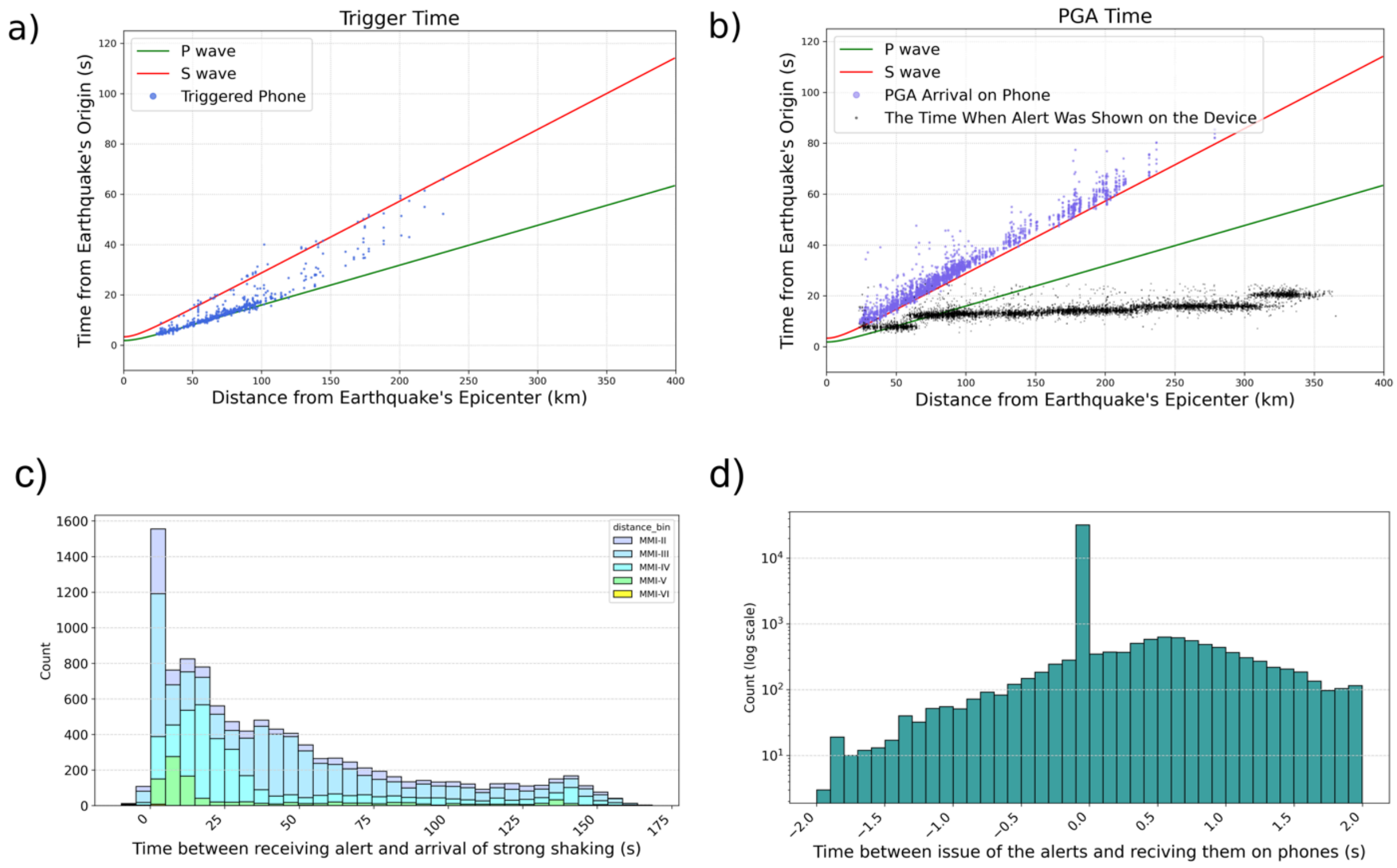

Figure 2: **Analysis of Alert Timing and Ground Motion Arrivals.** a) The triggering time of Android phones, with respect to the earthquake origin time, as a function of distance (top left) and b) arrival time of peak ground acceleration on phones (top right). The actual time when alerts were shown on phones are indicated by black dots. Theoretical arrival times of P and S waves are represented by solid green and red lines, respectively. The computed time between the arrival of alerts on phones and peak ground acceleration (c - bottom left) and the time between the issuing of the alerts and their delivery to the phones (d - bottom right) are also shown. The computed time between alert delivery and PGA (for all individual alerts) are color coded based on the epicentral distance of the phone and estimated average intensity.

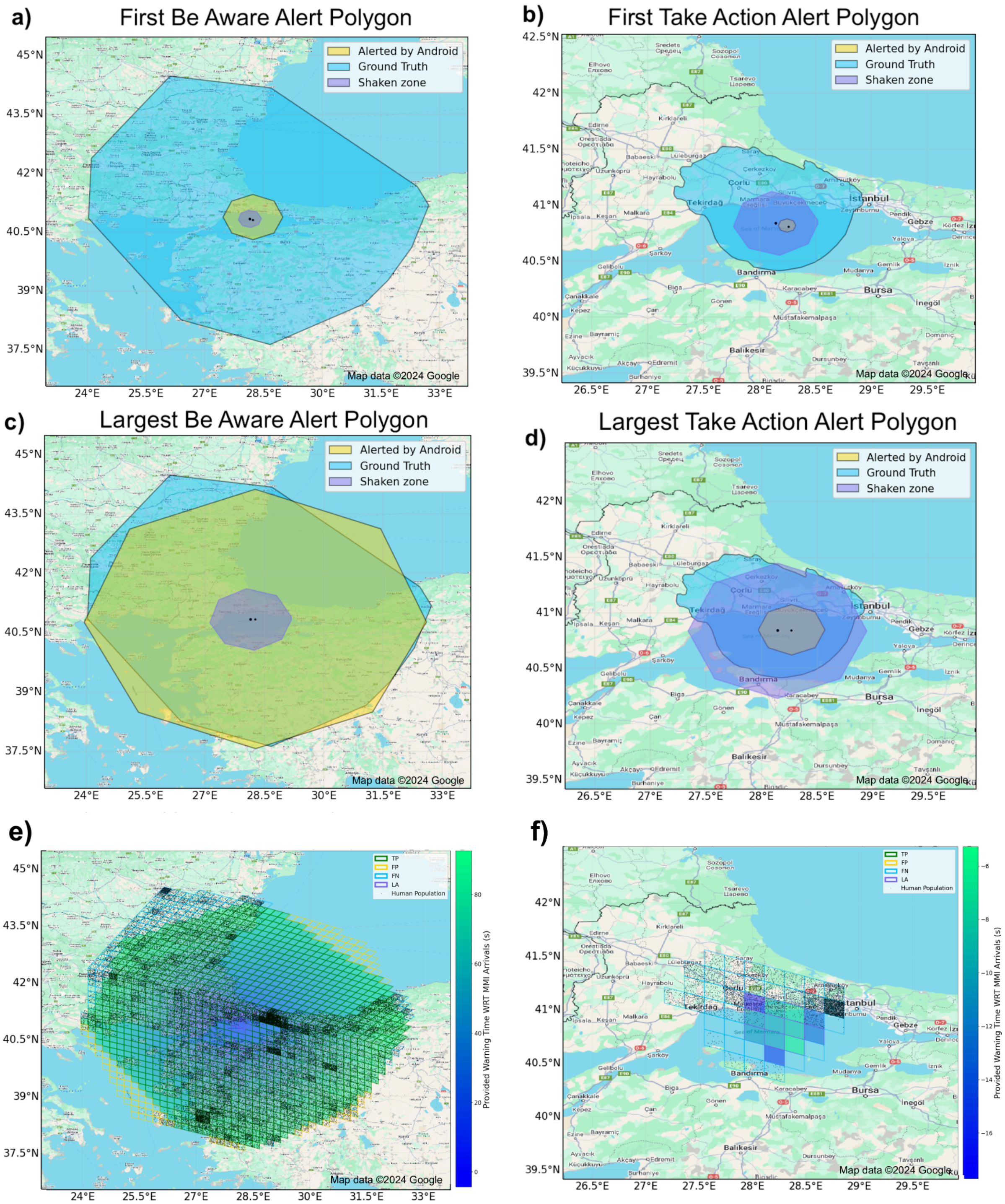

Figure 3: **Spatial Distribution and Performance Classification of AEA Alerts.** **(a)** First issued "Be Aware" alert polygon and **(b)** first issued "Take Action" alert polygon. **(c)** Largest issued "Be Aware" alert polygon and **(d)** largest issued "Take Action" alert polygon. AEA continuously refines magnitude estimates as seismic data arrives. Initial alerts are issued at $M_w > 4.5$, defining a circular notification

area based on magnitude and depth. Subsequent expansions occur as estimates increase (e.g., by 0.2 units), notifying all devices within the new boundaries. Panels (a–d) display the alerted regions, the final extent of MMI III shaking (ground truth), and the Shaken Zone representing wave propagation at the time of the alert. **(e)** Spatial grid classification for "Be Aware" alerts and **(f)** spatial grid classification for "Take Action" alerts, utilizing level 8 s2cells as the computational unit for utility metrics. Grids are color-coded by performance category: True Positive (TP; green), False Positive (FP; yellow), False Negative (FN; blue), and Late Alert (LA; purple). Black dots indicate population density. Correctly alerted cells (TP and LA) are further shaded to indicate warning times in seconds. Base maps generated using Cartopy; Map data ©2024 Google.

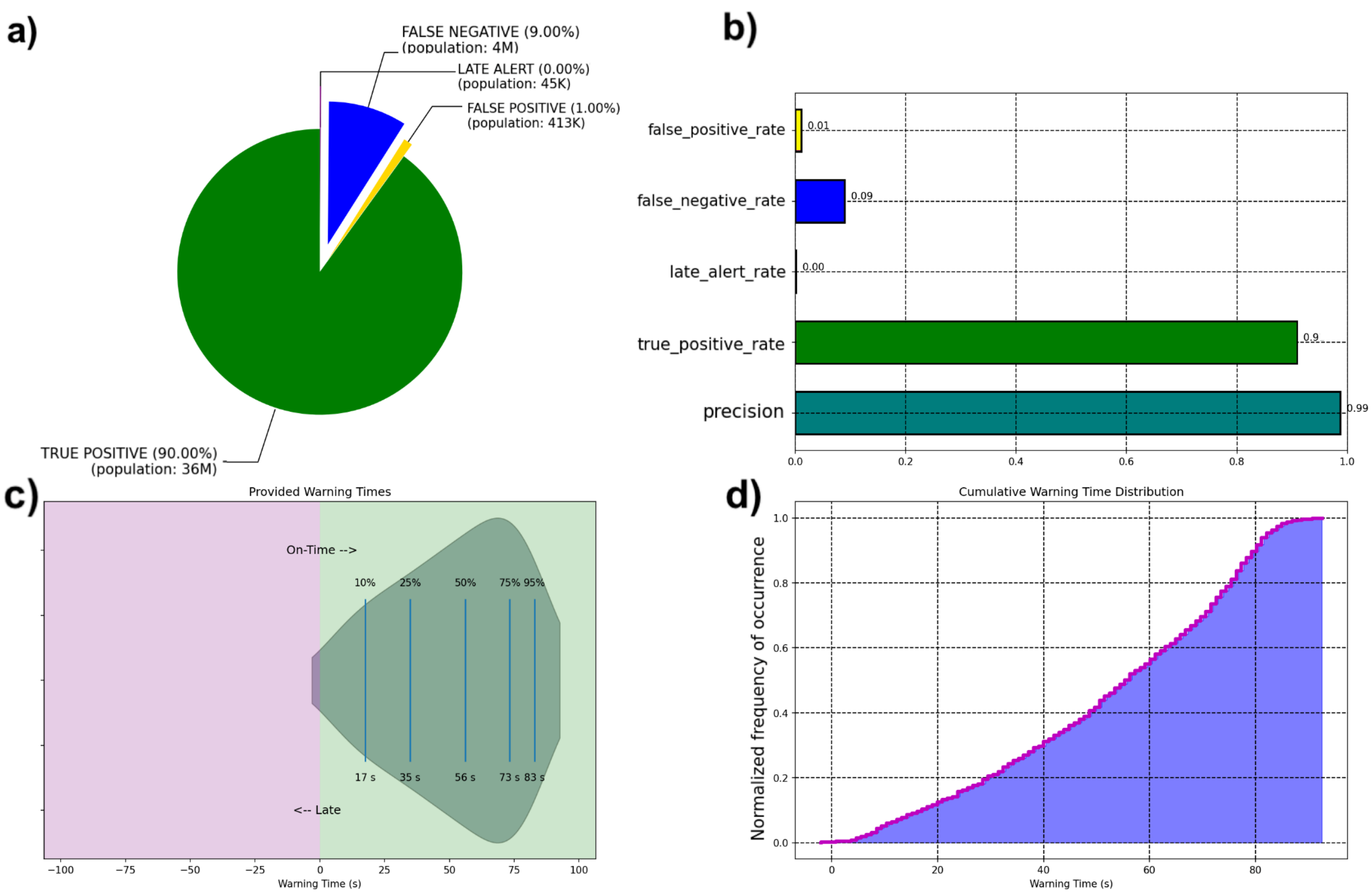

Figure 4: **Population Exposure and Warning Time Analysis for "Be Aware" Alerts.** a) The percentage of the total region's population (not just Android users) exposed to each alerting category. b) Alert performance metrics. c) The distribution of warning times provided to the "TRUE POSITIVE" population. The x-axis represents "Warning Time (s)". Negative values indicate the warning arrived *after* shaking started (late alerts), 0 means at the onset of shaking (S arrival), and positive values are seconds of warning before shaking. The y-axis implicitly represents the density or number of people receiving that warning time. d) This plot shows the cumulative frequency of warning times for the on-time alerted population.

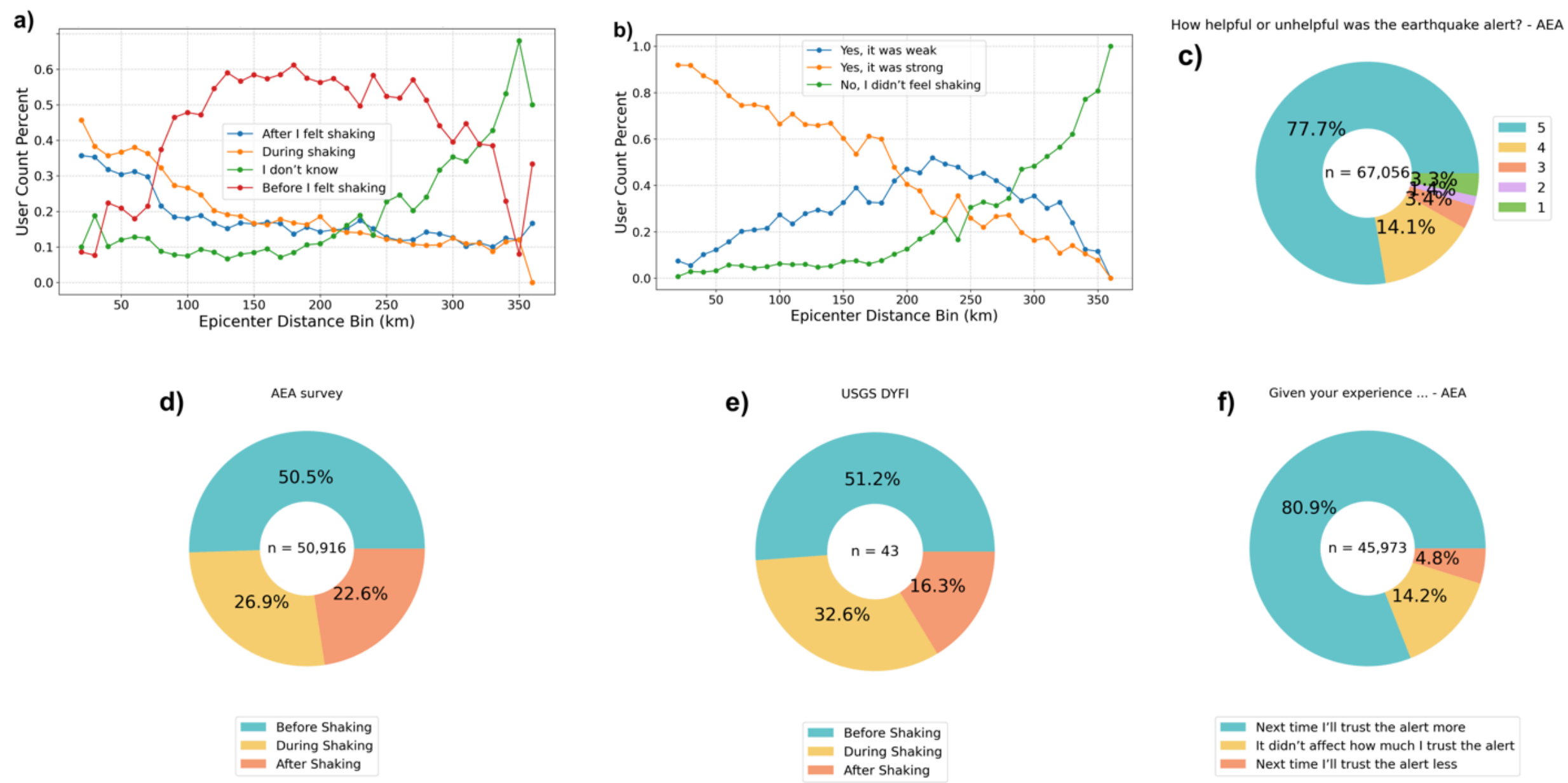

Figure 5: **Analysis of User Experience, Perceived Utility, and Trust from Feedback Surveys** (a) Distribution of responses to the question "When did you receive the alert?" as a function of distance to the epicenter. (b) Distribution of responses to the question "Did you feel shaking?" as a function of distance to the epicenter. (c) Distribution of responses to the question "How helpful or unhelpful was the earthquake alert?" in the AEA feedback survey. (d) Aggregate proportions of responses to the question "When did you receive the alert?" in the AEA feedback survey. (e) Distribution of responses to the question "Did you receive an alert message before or after you felt shaking?" in the independent USGS DYFI survey. (f) Distribution of responses to the prospective trust question ("Given your experience...") in the AEA feedback survey.

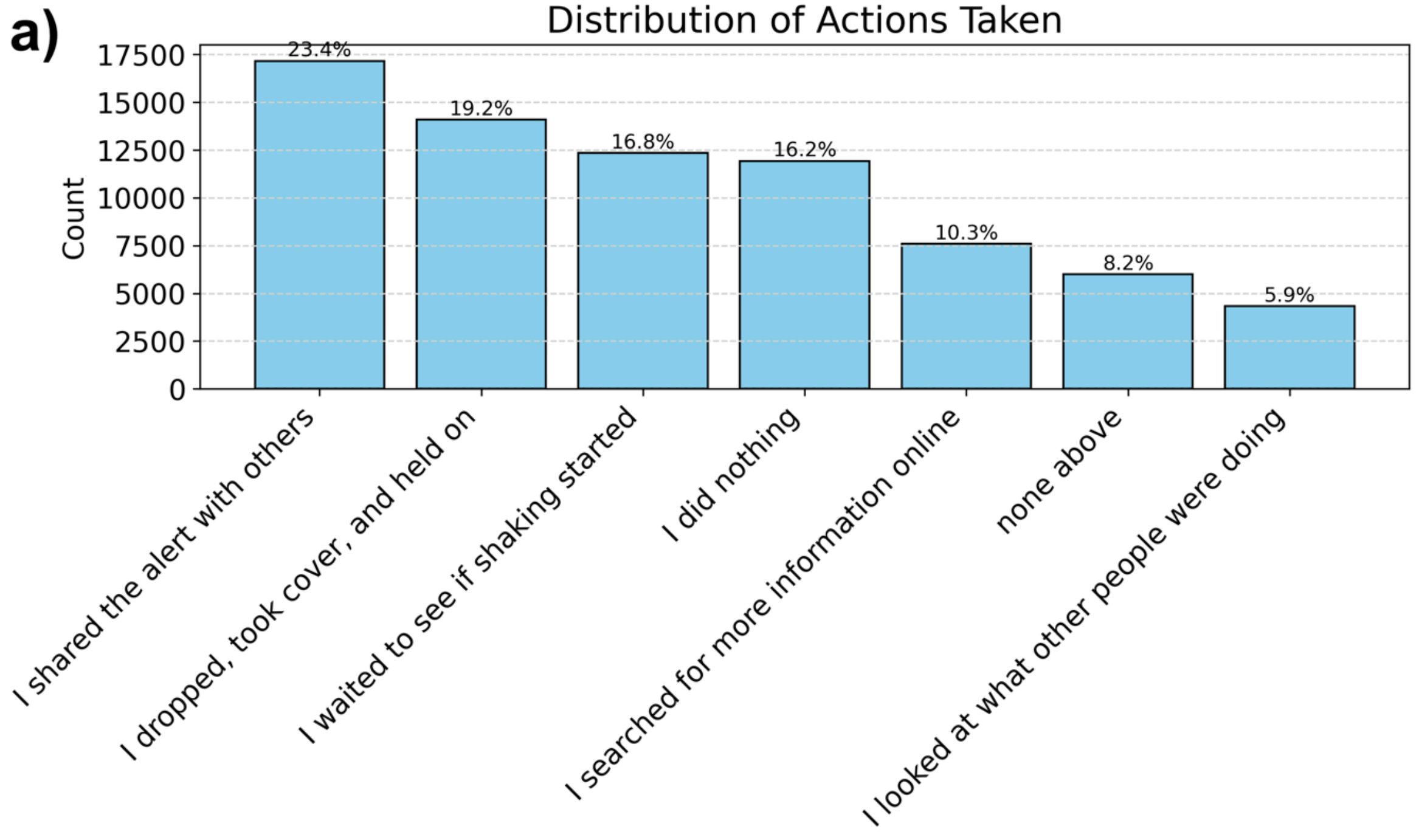

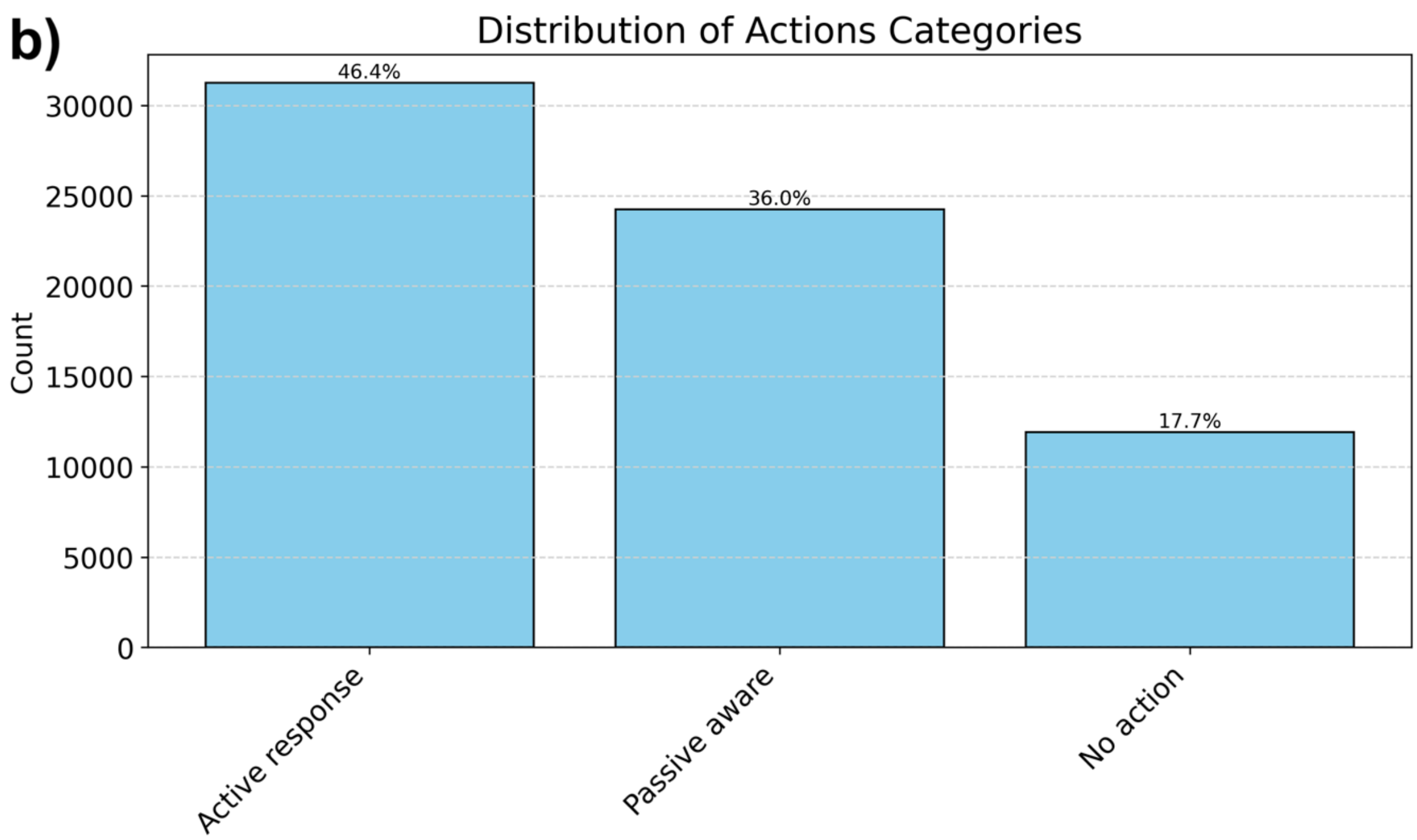

Figure 6: **User Actions and Behavioral Responses to AEA Alerts**. (a) Individual response frequencies to the question, "What did you do after receiving the alert?" in the AEA feedback survey ($N=67,433$). (b) The same behavioral data aggregated into three high-level categories of actions: active response, passive awareness, and inaction.

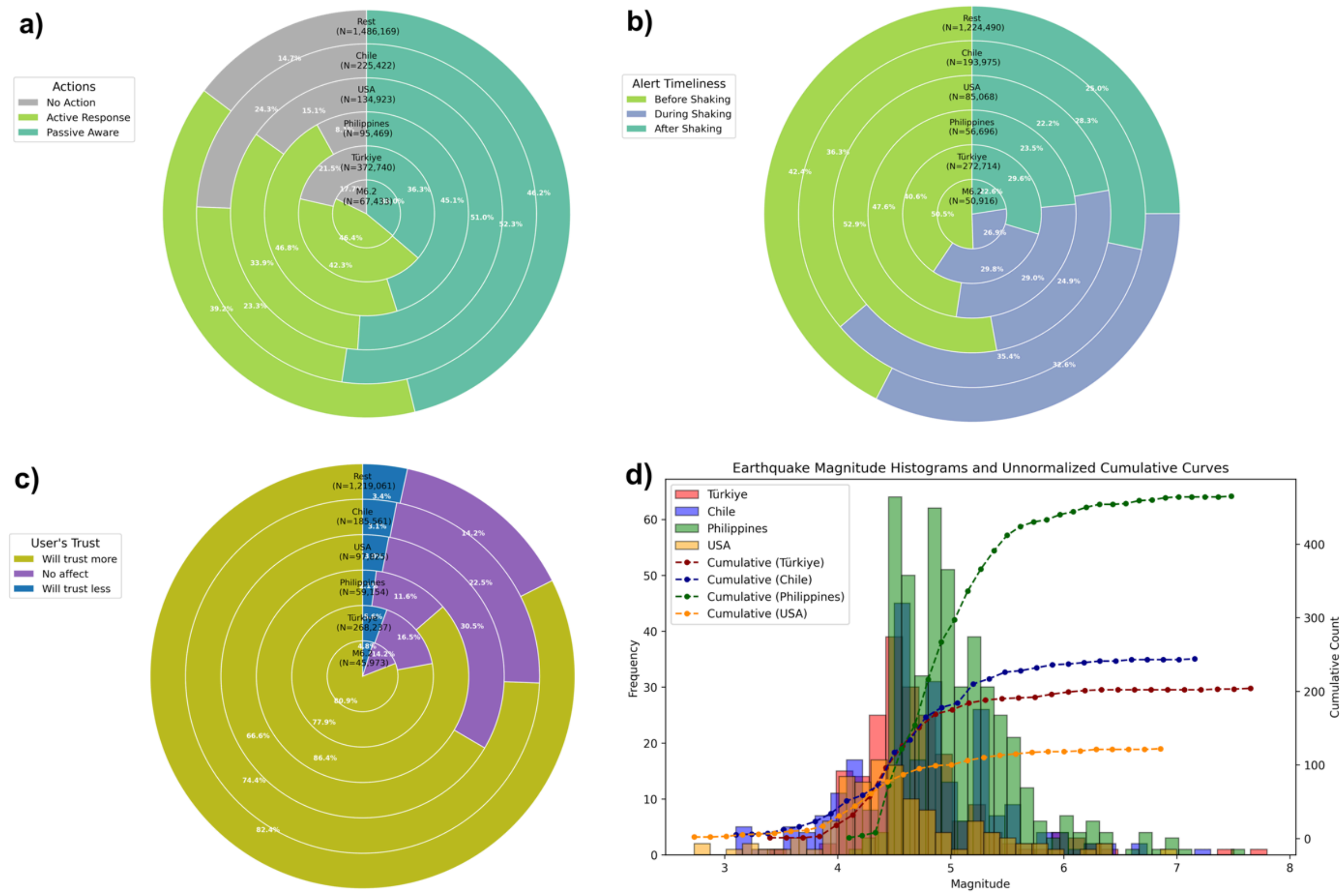

Figure 7: **Comparative Analysis of User Behavior, Alert Timeliness, and Trust across Geographic Regions**. (a) Comparison of human behavioral responses to AEA alerts across several geographic regions and the specific 2025 $M_w$ 6.2 Marmara earthquake. (b) Comparison of alert timeliness metrics across the same populations. (c) Comparison of changes in user trust across these cohorts. Each concentric ring represents a different population group, allowing for a direct comparison of their response patterns. (d) Distribution of historical earthquake magnitudes for Türkiye, Chile, the Philippines, and the USA. Bars represent the frequency of events (left axis), while dashed lines show the corresponding unnormalized cumulative count (right axis). Note that the alerted events in the US are based on the detection and estimates of USGS ShakeAlert, and AEA only delivers ShakeAlert alerts to Android users through its delivery mechanism.

Supplementary Materials for

# Performance and User Response of Android's Smartphone-Based Alerts in the 2025 Marmara Ereğlisi Earthquake

S. Mostafa Mousavi[1,2*], Patrick Robertson[3], Richard M. Allen[1,4], Alexei Barski[1], Robert Bosch[1], Nivetha Thiruverahan[1], Youngmin Cho[1], Tajinder Gadh[1], Steve Malkos[1], Boone Spooner[1], Greg Wimpey[1], Edward Shi[1], Marc Stogaitis[1]

[1] Google LLC, Mountain View, CA, USA.

[2] Department of Earth and Planetary Sciences, Harvard University, Cambridge, MA, USA.

[3] Google Germany GmbH, Munich, Germany.

[4] Seismological Laboratory, University of California, Berkeley, Berkeley, CA, USA.

*Corresponding author. Email: mousavim@google.com

**The PDF file includes:**

Table S1, S2
Figures. S1 to S6

Table S1: Real-time production system detection of April 23, 2025, M6.2 Marmara Ereğlisi (EMSC Event ID: 20250423_0000104) earthquake,  Türkiye.

| Time with respect to EMSC origin time of 09:49:11.9 UTC (sec) | Estimated magnitude [EMSC Mw 6.2] | Estimated epicenter location (latitude, longitude) | Location error relative to EMSC location of 40.833°N 28.227°E (km) | Estimated depth (km) [EMSC estimated depth 15.0 km] | BeAware, TakeAction alert distances from ground truth epicenter (km) |
|---|---|---|---|---|---|
| 5.31 s | 4.83 | 40.88, 28.229 | 5.0 km | -5 km | 70.00, 0.0 |
| 8.94 s | 5.03 | 40.818, 28.247 | 2.0 km | -6 km | 91.82, 10.34 |
| 9.87 s | 5.32 | 40.82, 28.252 | 3.0 km | 0 km | 163.26, 11.92 |
| 10.89 s | 5.58 | 40.826, 28.252 | 2.0 km | 8 km | 233.90, 14.16 |
| 12.71 s | 5.82 | 40.833, 28.262 | 3.0 km | 18 km | 329.31, 28.23 |
| 17.12 s | 6.04 | 40.846, 28.272 | 4.0 km | 32 km | 365.08, 29.03 |
| 24.11 s | 5.82 | 40.846, 28.277 | 4.0 km | 12 km | 330.30, 37.33 |
| 32.56 s | 5.62 | 40.838, 28.267 | 3.0 km | -3 km | 281.15, 34.02 |

Table S2: M6.2 event and its immediate aftershocks.

| Event date-time | Magnitude (source) | AEA Detection Time | AEA Max Magnitude |
|---|---|---|---|
| 2025-04-23 09:49:11 (UTC) | M6.2 (USGS) | 09:49:17.21 | M6.04 |
| 2025-04-23 09:51:19 (UTC) | M4.7 (KOER), M5.3 (EMSC), M5.9 (AFAD) | 09:51:27.31 | M4.13 |
| 2025-04-23 10:02:02 (UTC) | M4.5 (AFAD) | 10:02:16.98 | M3.8 |
| 2025-04-23 10:02:32 (UTC) | M5.2 (KOER), M5.0 (EMSC), M4.9 (AFAD) | – | – |

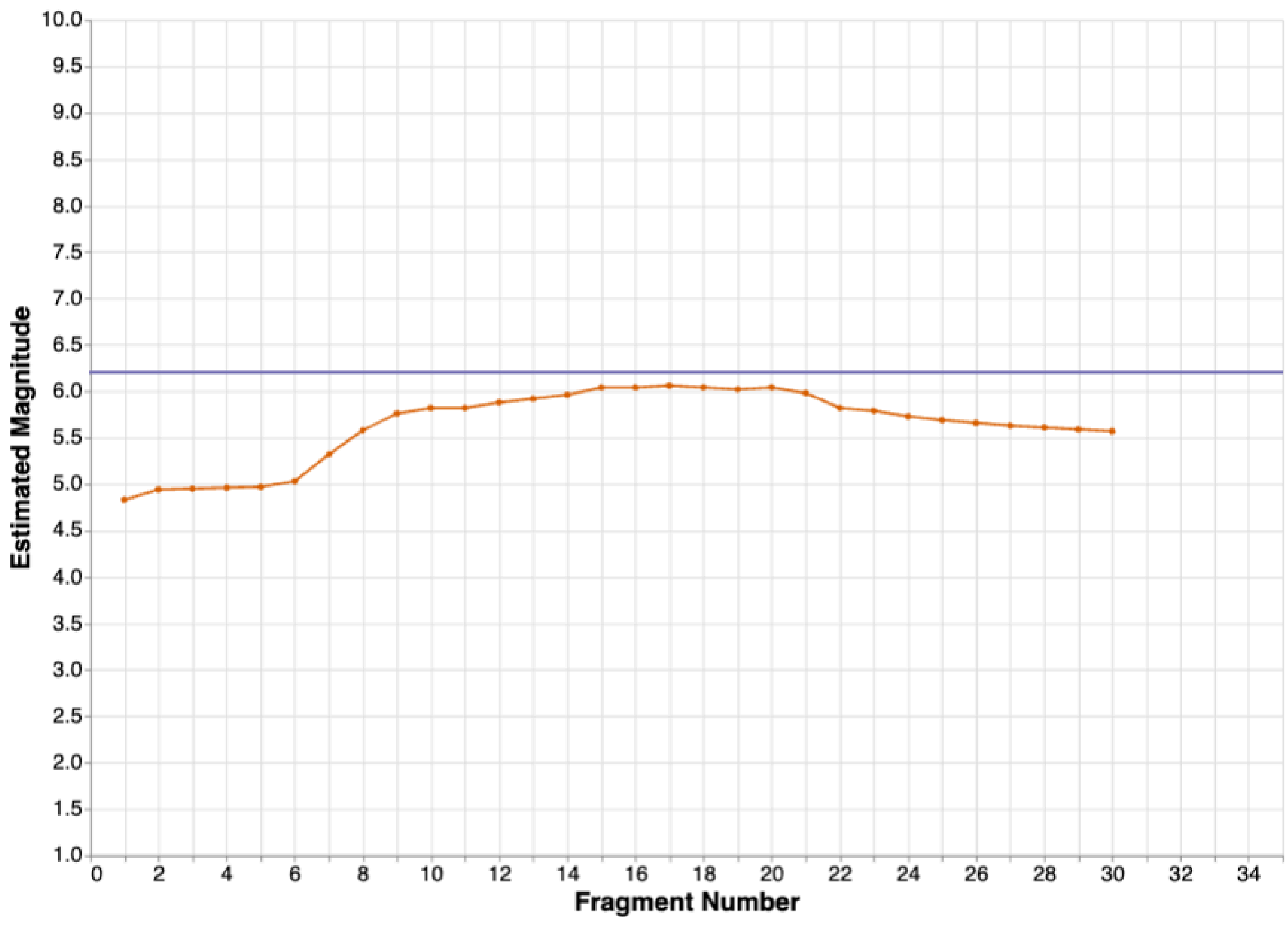

Figure S1: **Real-time Magnitude Estimation Evolution Over Time.** The plot illustrates the evolution of estimated magnitudes for the M6.2 Marmara Ereğlisi earthquake by AEA (red line) compared against the ground truth magnitude, which represents the best final estimate from the USGS (horizontal blue line). AEA estimates the earthquake's source parameters at 1.0-second

time intervals based on new data arrivals, with each individual orange dot representing a system calculation using all aggregated data collected up to that time stamp.

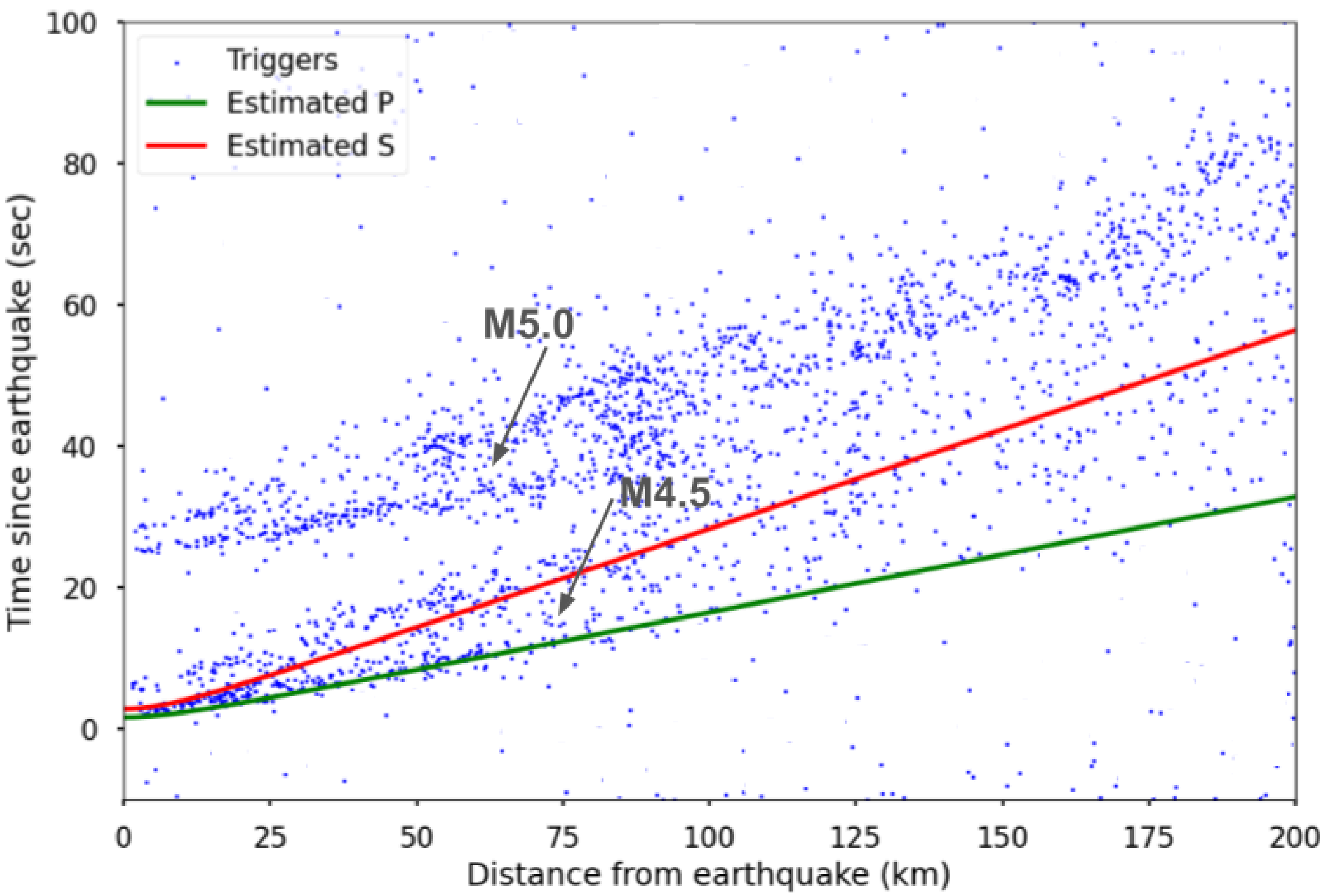

Figure S2: **Spatial Triggering Patterns for Closely Spaced Aftershocks.** This figure maps the triggered phones recorded during the M4.5 (23-04-2025 10:02:10 UTC) and M5.0 (23-04-2025 10:02:32 UTC) seismic events. The first event was detected by AEA at 10:02:17 UTC as a Mw 3.8 aftershock fitting the P- and S-wave arrival windows, whereas the second event arriving approximately 25 seconds later went undetected as a separate alert due to its close spatial and temporal proximity to the first.

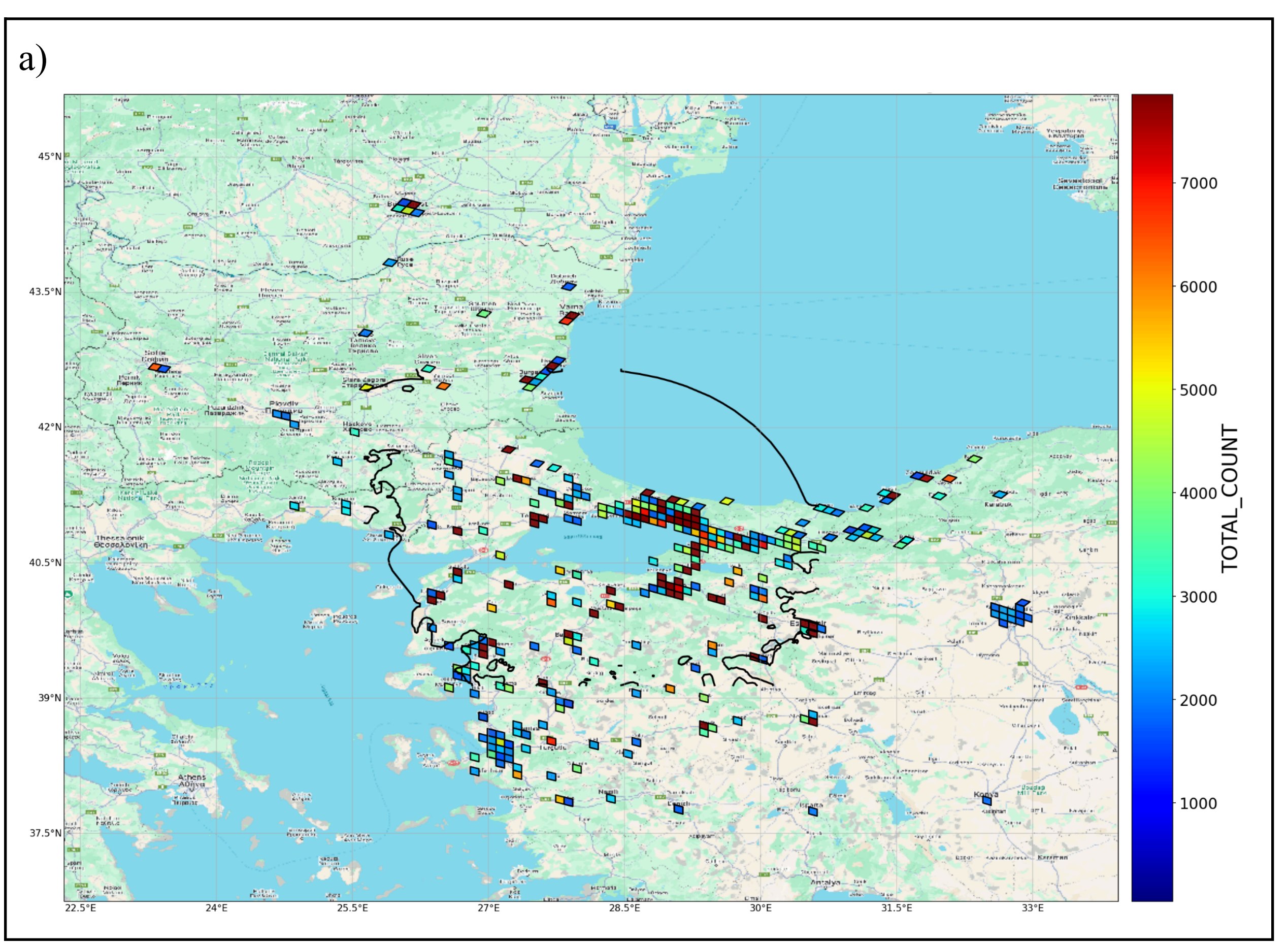
a)
45°N
43.5°N
42°N
40.5°N
39°N
37.5°N
22.5°E
24°E
25.5°E
27°E
28.5°E
30°E
31.5°E
33°E
7000
6000
5000
4000
3000
2000
1000
TOTAL_COUNT

b)

45°N
43.5°N
42°N
40.5°N
39°N
37.5°N

22.5°E 24°E 25.5°E 27°E 28.5°E 30°E 31.5°E 33°E

YES_RATIO
1.0
0.8
0.6
0.4
0.2

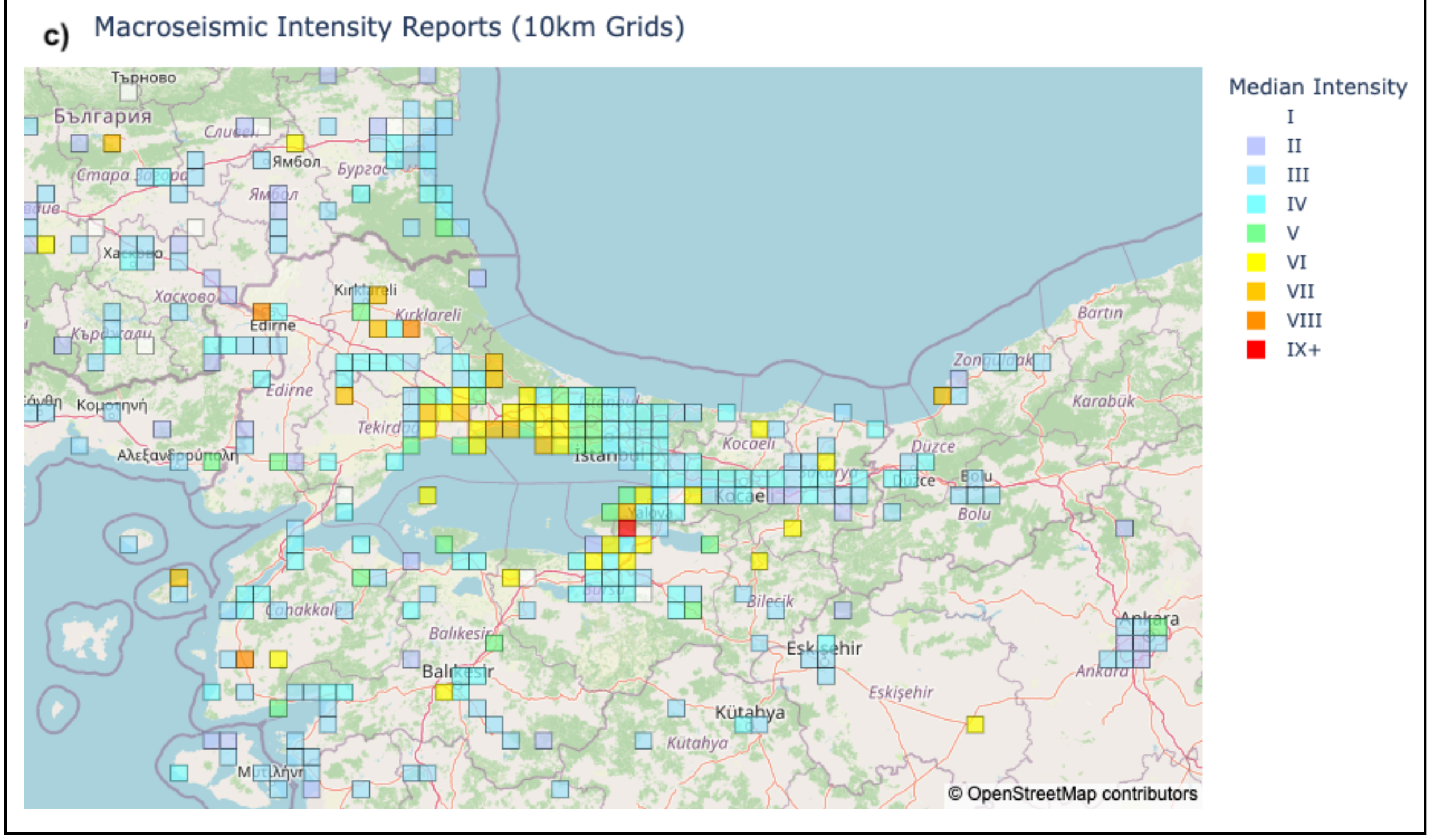

Figure S3: **Search Survey Proportions and Macro-Seismic Felt Reports.** The grid layouts illustrate the total count of search survey responses (panel a) and the ratio of affirmative "yes" responses to total responses (panel b) across 10-km-wide geographic bins. The overlaid black line shows the INGV MMI III ground-shaking intensity contour. Panel c displays the distribution of around 5,000 aggregated (into a 10 km × 10 km spatial grids) felt reports collected by the USGS DYFI system and EMSC. The color of each grid cell represents the median Community Decimal Intensity (CDI) reported within that cell, categorized into Modified Mercalli Intensity (MMI) classes (I through IX+). Colors correspond to the standard USGS intensity color palette. a) and b) map subplots are generated using Python library with basemaps from Google Maps Web Map Tile Service (accessed via Cartopy). Map data ©2024 Google. c) was created using the PlotLy python library and an Open Street style base map. Map data © OpenStreetMap contributors. OpenStreetMap content is published under the Open Database License (ODbL), and map tiles/content are licensed under the Creative Commons Attribution-ShareAlike 2.0 (CC BY-SA 2.0) License. Licensing and copyright details can be found at https://www.openstreetmap.org/copyright and https://creativecommons.org/licenses/by-sa/2.0/. No pre-existing graphical templates, stock icons, or proprietary third-party design elements were used. All components represent raw, author-generated telemetry plots, survey distributions, and analytical spatial data rendered directly from source code.

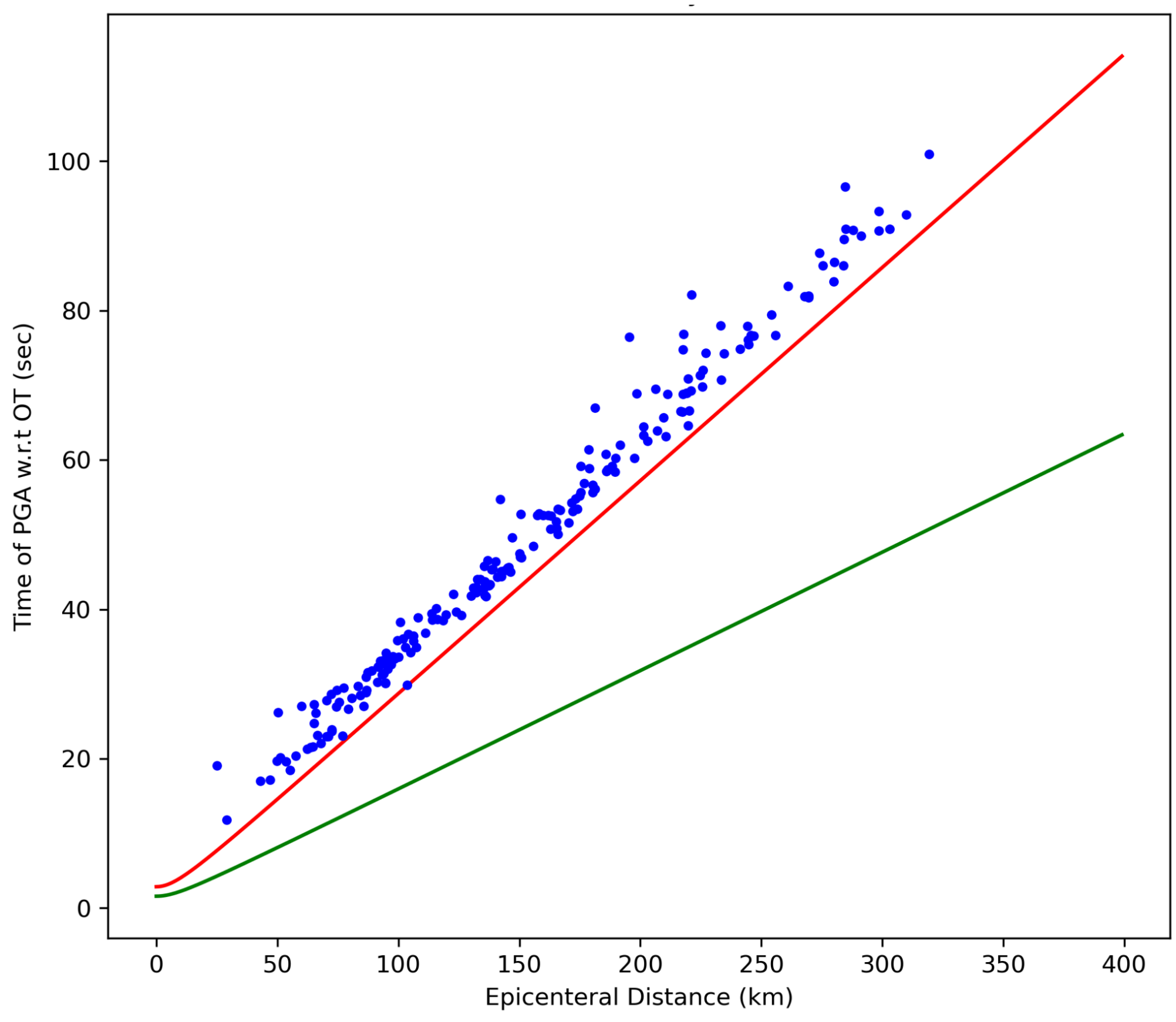

Figure S4: **Observed Peak Ground Acceleration Arrival Times Versus Distance.** The scatter plot delineates the arrival times of peak ground acceleration (PGA) recorded across the regional AFAD seismic network for the April 23, 2025, $M_w$ 6.2 Marmara Ereğlisi earthquake. Individual blue dots represent the observed time of PGA at specific seismic stations relative to the origin time, which are plotted alongside solid green and red curves that represent the theoretical arrival times for P-waves and S-waves, respectively.

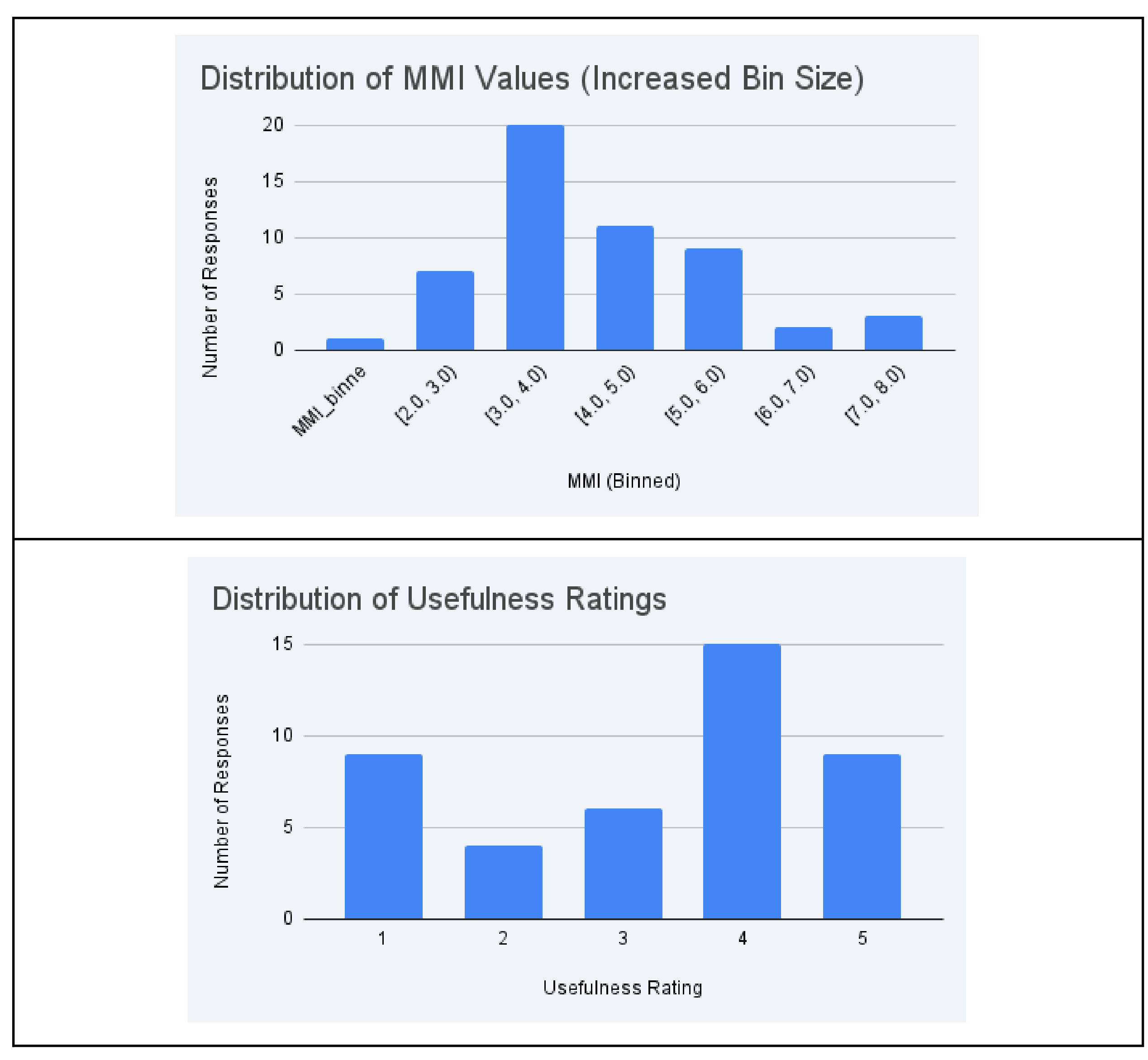

Figure S5: **Supplementary Questionnaire Intensity Distributions and Helpfulness Ratings.** The charts summarize citizen scientist experiences by showing the DYFI intensity estimates associated with the 55 responders to the USGS supplementary questionnaires (top panel) alongside the corresponding distribution of usefulness ratings (bottom panel). These independent survey metrics provide macroseismic calibration data by correlating the user-perceived utility of the early warning notification directly against the estimated Modified Mercalli Intensity of local ground shaking.

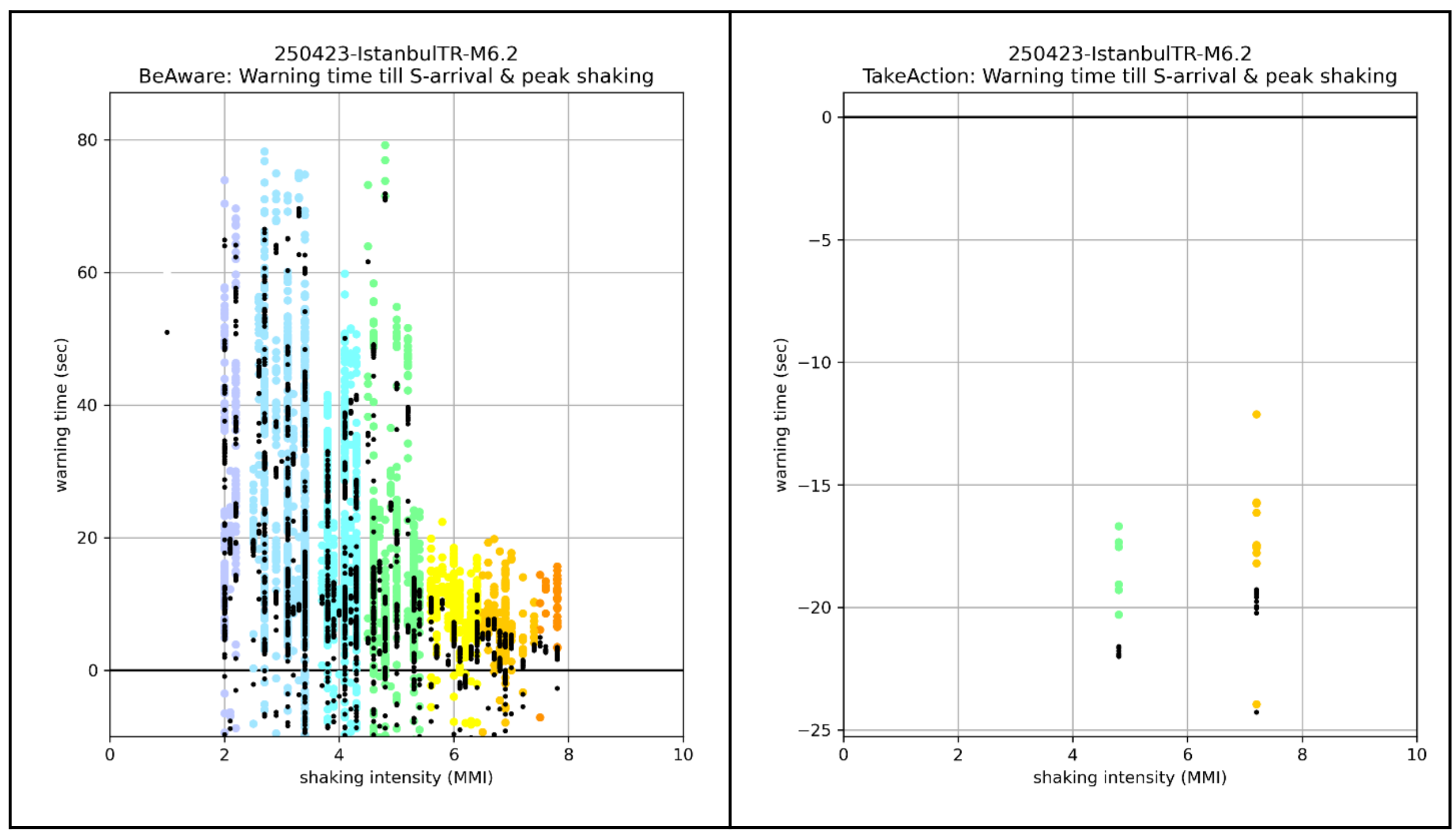

Figure S6: **Delivered Warning Lead Time As a Function of Shaking Intensity.** The scatter plots compare the BeAware (left panel) and TakeAction (right panel) alert warning times plotted against the closest corresponding USGS DYFI macroseismic report within a 5-km radius. Black dots indicate the available lead time measured from alert delivery on a handset until the nominal S-wave arrival, whereas the colored dots represent the warning time remaining until the true peak shaking was registered on the device, color-coded according to the standard ShakeMap intensity scale.

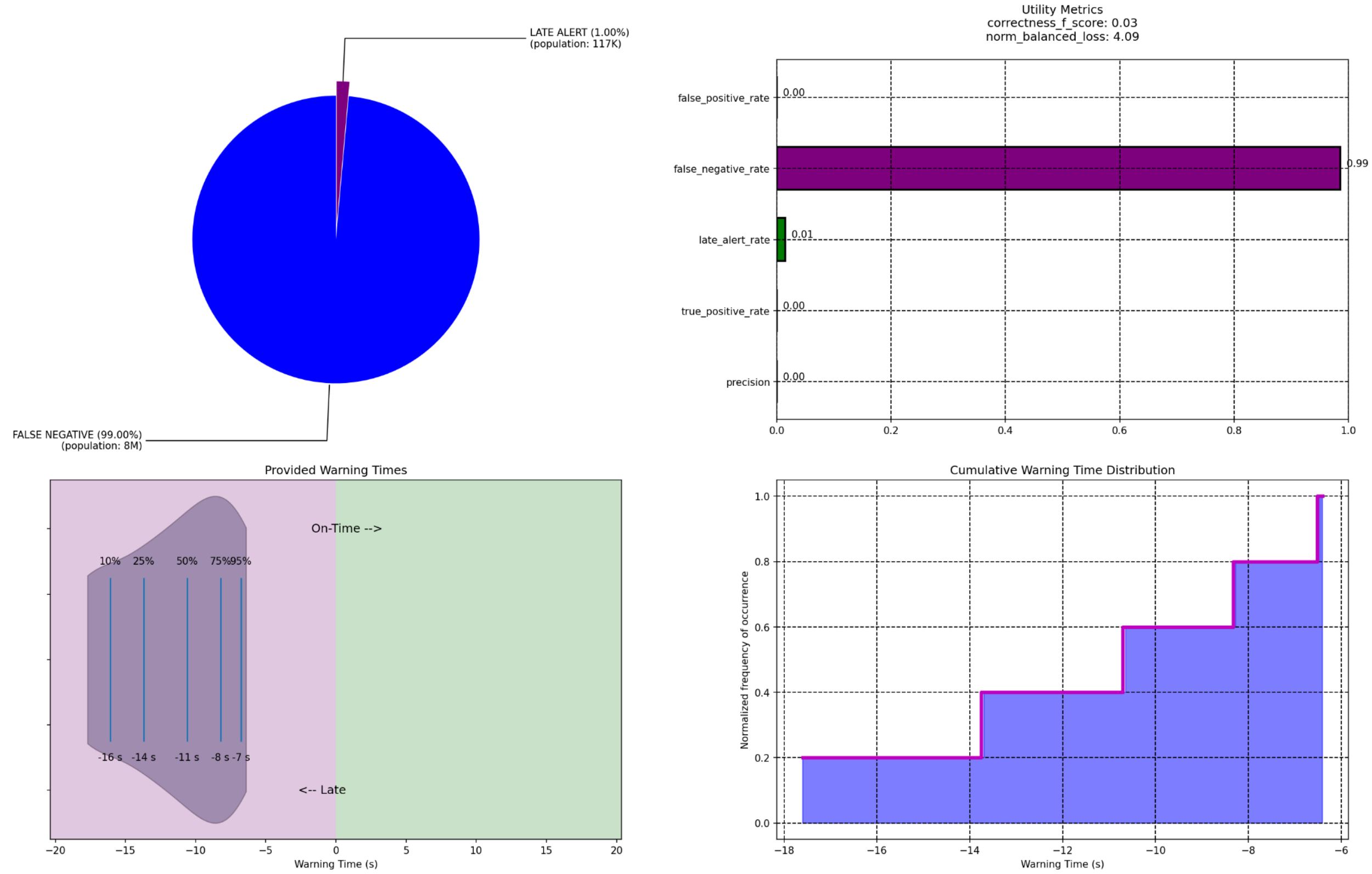

Figure S7: **Performance Evaluation Metrics and Warning Time Cumulative Frequencies.** The multi-panel figure provides a comprehensive performance evaluation of the TakeAction alerting tier across the general human population exposed to the earthquake. The layout illustrates the percentage of the total regional population exposed to each alerting category (panel a), key statistical performance metrics (panel b), the probability density of warning times provided to the true positive population (panel c), and the cumulative frequency distribution of warning times for on-time alerted users (panel d).